\documentclass[%
reprint,
superscriptaddress,
floatfix,
amsmath,
amssymb,
aps,
notitlepage
]{revtex4-1}
\usepackage[hidelinks=true]{hyperref}
\usepackage{xcolor}
\definecolor{greenish}{rgb}{0.13,0.58,0.16}
\definecolor{reddish}{RGB}{174,12,48}
\definecolor{blueish}{rgb}{0.12, 0.56, 1.0}
\definecolor{magenta}{rgb}{0.8, 0.0, 0.8}

\hypersetup{
colorlinks   = true, 
urlcolor     = greenish, 
linkcolor    = magenta, 
citecolor   = blueish 
}

\usepackage{setspace}
\usepackage{tikz,pgfplots}
\usetikzlibrary{external}
\pgfplotsset{
	compat=newest,
	invoke before crossref tikzpicture={\tikzexternaldisable},
	invoke after crossref tikzpicture={\tikzexternalenable},
}
\usetikzlibrary{
	external,
}
\pgfplotsset{plot coordinates/math parser=false}
\usetikzlibrary{plotmarks}
\usepackage{caption}
\usepackage{siunitx}
\usepackage{subcaption}
\usepackage{graphicx}
\usepackage{dcolumn}
\usepackage{bm}

\def\el{\text{el}}
\def\trans{\text{trans}}
\def\axial{\text{axial}}
\def\phi{\varphi}
\def\xB{\mathbf{x}}
\def\x{\text{x}}
\def\y{\text{y}}
\def\E{\mathcal{E}}

\def\fix{\text{fix}}

\def\bnd{\text{bnd}}
\def\R{\mathbb{R}}
\def\K{\mathcal{K}}

\def\tr{\text{tr}}

\def\A{\mathbb{A}}
\def\aB{\mathbf{a}}
\def\aBt{\tilde{\mathbf{a}}}
\def\bB{\mathbf{b}}
\def\bBt{\tilde{\mathbf{b}}}
\def\T{\mathbf{T}}
\def\P{\mathbf{P}}
\def\Rbd{\mathbf{R}}
\def\v{\mathbf{v}}
\def\r{\mathbf{r}}
\def\N{\mathbf{N}}
\def\d{\text{d}}
\def\K{\mathcal{K}}

\def\a{\mathbf{a}}
\def\b{\mathbf{b}}
\def\const{\text{const.}}

\def\al{\bm{\alpha}}
\def\be{\bm{\beta}}
\def\ta{\tilde{a}}
\def\tb{\tilde{b}}
\def\th{\hat{\mathbf{t}}}
\def\bh{\hat{\bm{\zeta}}}
\def\dh{\hat{\bm{\eta}}}
\def\O{\mathcal{O}}

\newcommand{\blue}[1]{\textcolor{black}{#1}}

\begin{document}
	\title{Totimorphic assemblies from neutrally-stable units}
	\author{Gaurav Chaudhary}
	\thanks{Equal contribution}
	\affiliation{School of Engineering and Applied Sciences, Harvard University, Cambridge, MA 02138.}
	\author{S Ganga Prasath}
	\thanks{Equal contribution}	
	\affiliation{School of Engineering and Applied Sciences, Harvard University, Cambridge, MA 02138.}
	\author{Edward Soucy}
	\affiliation{Center for Brain Science, Harvard University, Cambridge, MA 02138.}
	\author{L Mahadevan}
	\email{lmahadev@g.harvard.edu}
	\affiliation{School of Engineering and Applied Sciences, Harvard University, Cambridge, MA 02138.}
	\affiliation{Center for Brain Science, Harvard University, Cambridge, MA 02138.}
	\affiliation{Department of Physics, Harvard University, Cambridge, MA 02138.}
	\affiliation{Department of Organismic and Evolutionary Biology, Harvard University, Cambridge, MA 02138.}


	\begin{abstract}
Inspired by the quest for shape-shifting structures in a range of applications, we show how to create morphable structural materials using a neutrally stable unit cell as a building block. This unit cell is a self-stressed hinged structure with a one-parameter family of morphing motions that are all energetically equivalent; however, unlike kinematic mechanisms, it is not infinitely floppy and instead exhibits a tunable mechanical response akin to that of an ideal rigid-plastic material. Theory and simulations allow us to explore the properties of planar and spatial assemblies of neutrally-stable elements and also pose and solve the inverse problem of designing assemblies that can morph from one given shape into another. Simple experimental prototypes of these assemblies corroborate our theoretical results and show that the addition of switchable hinges allows us to create load-bearing structures. All together, totimorphs pave the way for structural materials whose geometry and deformation response can be controlled independently and at multiple scales.
\end{abstract}

	\pacs{Valid PACS appear here}
	\maketitle

	Arange of mechanical systems that arise in soft robotics, flexible antennae, prosthetics etc.  require transitions between multiple relatively stable conformational states that are all easily accessible without large energetic barriers between them  ~\cite{rus2015design,whitesides2018soft}. Recent advances in manufacturing technologies are a step in this direction and have enabled the fabrication of shape morphing structures that respond to a range of external stimuli such as heat, light, humidity, pH, magnetic or electric fields etc.~\cite{gladman2016biomimetic,boley2019shape}. However, these structures typically have only a finite number of metastable equilibrium states separated from each other by energetic barriers that can be overcome only in the presence of a guided external energy input. This raises a natural question: can one design a material structure that is so malleable that it has an infinite number of energetically equivalent states? And can one control the possibly heterogeneous response of these structures on multiple scales? A possible set of candidates are mechanisms, floppy structures that are akin to a set of underconstrained linkages that have a number of zero-energy internal modes. However, such  systems are unstable to small perturbations and difficult to control. To create mechanism-like infinitely malleable (totimorphic) structures but are also stable and controllable, we need to have a finite resistance to deformation that is independent of the magnitude of deformation. This study provides one possible answer suggested by the use of neutrally stable structural assemblies that are totimorphic, stable and controllable.

Our starting point is the observation of Maxwell~\cite{maxwell1864calculation} who defined the criteria associated with  stable mechanical structures in terms of the difference between the number of degrees of freedom and the number of constraints in the system. When the difference is exactly zero, the structure is isostatic and stable, while when the difference is either positive or negative, we get either under-constrained structures with zero energy \textit{floppy modes}, or over-constrained systems with states of self-stress \cite{calladine1978buckminster}. An unusual class of structures that sits precariously at the boundary between isostatic and floppy structures are neutrally stable (NS) structures ~\cite{schenk2014zero},  marginally stable objects typically made of a combination of very stiff and very soft elements that are internally strained but globally equilibrated. A key feature of these structures is the presence of geometric constraints  that allow for a one (or more) parameter family of large deformations that redistribute the internal stresses while keeping the total strain energy invariant. Thus, no additional external work is required to change or maintain their configurations making these objects infinitely malleable. This makes neutrally stable structures (NSS) different from typical monostable or even multistable unstrained or prestrained (e.g tensegrity~\cite{tibert2003review}) structures which require external work to deform them to switch between states. Instead, NSSs exhibit an infinity of equivalent equilibrium states that are locally stabilized by internal friction and require minimal external work to transition between them.

An everyday example of a NSS is the Anglepoise lamp~\cite{french2000spring,george1937equipoising} designed almost a century ago, with a lamp head that is infinitely morphable by virtue of its having a set of opposing springs in tension that change their lengths while the total energy remains constant. 
The presence of finite frictional or other dissipative forces provide local stability to the local conformation of the NSS, but this is not a robust approach for assemblies of NSSs. To use these as the basis for creating more complex shapes, we need to $(i)$ develop a generalized approach to the design and assembly of NSS, $(ii)$ develop methods to ``lock'' and ``unlock'' the NSS once it has reached a pre-specified conformation. Here we describe an approach to develop neutral stability based totimorphs that are geometrically complex, and can be mechanically stabilized or destabilized structurally on-demand.

\begin{figure*}[htb!]
\centering
\includegraphics[width=0.95\textwidth]{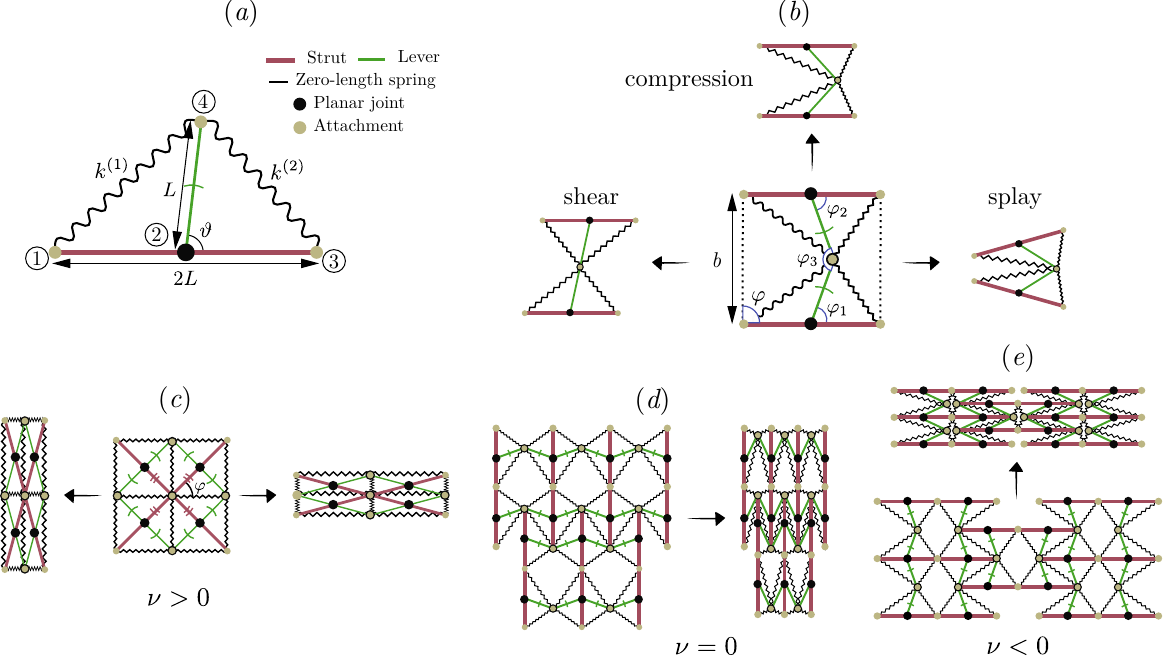}
\caption{\textbf{Neutrally-stable unit cells and their assembly}. $(a)$ Schematic of a neutrally-stable (NS) unit cell constructed from two rigid elements, a strut and a lever, and two elastic elements associated with zero natural-length springs of stiffness $k^{(1)} = k^{(2)}$.   $(b)$ A simple planar assembly of two unit cells allow for three deformation modes (shear, splay and compression). Combination of such modes enable the deformation of NS assemblies (with multiple unit cells) to various configurations.  $(c-e)$ Assembly of NS structures can have Poisson's ratio  defined as $\nu = -{\varepsilon_\text{trans}}/{\varepsilon_\text{axial}} $, that is $(c)$ positive  with $\nu = \cot^2 \phi$, $(d)$ zero , and $(e)$ negative (auxetic)  with $\nu = -{\cot \phi}/{\cot (\phi/2)}$. Multiple struts are rigidly connected (in $(d, e)$) or through a pin joint (in $(c)$). In all diagrams, a yellow-gray disk indicates the attachment of a spring to a rigid element, a black disk denotes connections between rigid elements via pin joints,  a yellow-gray disk surrounded by a black circle indicates a pin joint between a pair of rigid elements, and arcs on levers represent independent rotational degrees of freedom.}
\label{fig:fig1}
\end{figure*}

Our structures use a minimal NS unit cell as the basis for our assembly shown schematically in Fig.~\ref{fig:fig1}$(a)$ and physically in Fig.~\ref{fig:fig3}$(a)$. It is constructed using two ``zero-length'' springs~\cite{delissen2017design} (i.e. springs that are stretched substantially relative to their rest length) (denoted by $S$), a lever of length $L$, and a rigid link of length $2L$ (springs between lever to strut are connected through point attachments). The freely rotating lever is attached to the link via a planar joint (with 1 DoF) and allows for a range of configurations characterized by a single internal angle $\vartheta$. For each value of this angle, the individual springs with stiffness $k^{(1)}, k^{(2)}$ are stretched differently, and the total energy of this system is given by
\begin{align*}
  \E_{\el}&  = \ \frac{1}{2} k^{(1)} (l^{(1)})^2 + \frac{1}{2} k^{(2)} (l^{(2)})^2\\
  & = \ 2L^2 \bigg[ k^{(1)} \cos^2 \bigg(\frac{\vartheta}{2}\bigg) + k^{(2)} \sin^2 \bigg(\frac{\vartheta}{2}\bigg) \bigg]
\end{align*}

For the special case $k^{(1)} = k^{(2)}$ (which we will restrict ourselves to - but see SI for details of how this is not as restrictive as it seems), this energy is independent of $\vartheta$, and the structure has a constant total energy for all orientations of the lever - this can be viewed as a mechanical realization of Thales theorem.

\begin{figure*}[!]
\centering
\includegraphics[width=0.75\textwidth]{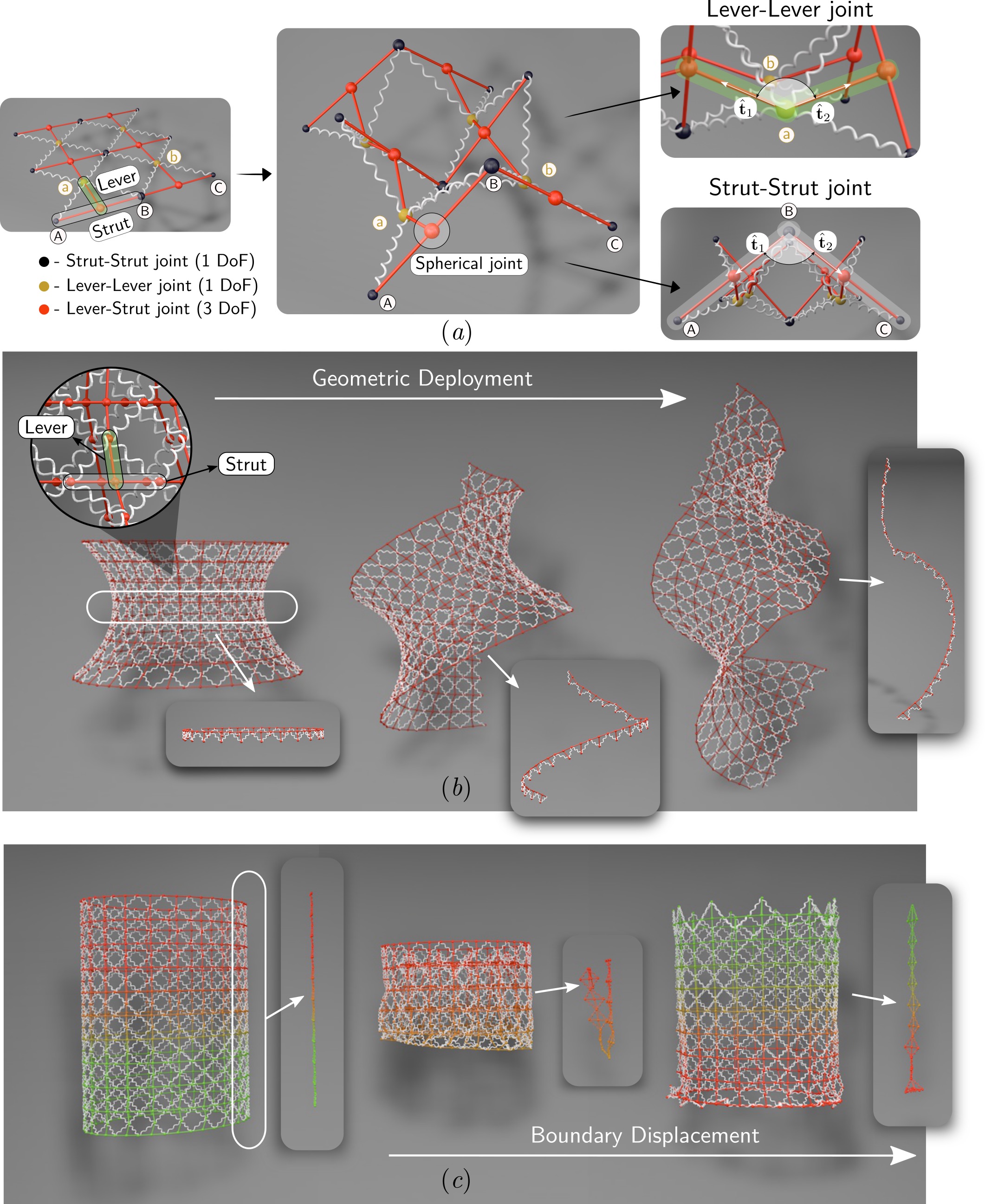}
\caption{\textbf{Simulations of totimorphic assemblies using neutrally-stable units.} $(a)$ A NS-net assembled from NS unit cells can undergo out-of-plane deformation. The lever and the strut in the unit cell are connected via a spherical joint, and the connections between the neighboring cells are made by NS joints (same DoF as a pin joint, see Fig.~S7 for details). The effective motion because of such connections is highlighted. $(b)$ Deployment of a catenoid, tessellated using with 342 NS unit cells (connections highlighted in the zoomed image). The catenoid morphs into a helicoid via geometric mapping of the surface (see SI sec.~S5). Inset shows the large displacement of the highlighted section of the catenoid. $(c)$ Eversion of an open cylinder tessellated from 420 unit cells. The top of the cylinder is fixed, while the bottom end is first compressed and then displaced vertically in small steps until the entire structure is turned inside-out. All the intermediate states during the eversion are mechanically stable. Inset highlights the folding of an axial strip.}
\label{fig:fig2}
\end{figure*}
To extend the capability of the minimal unit cell shown in Fig.~\ref{fig:fig1}$(a)$, we construct a simple planar NSS by pinning the levers of two NS unit cells as shown in Fig.~\ref{fig:fig1}$(b)$. Using the Tchebychev-Gr\"ubler-Kutzback (CGK) criteria in 2D \cite{marmier2016flexibility}, we can calculate the number of degrees of freedom (DoF) of such a structure with $n$ rigid bodies connected at $g$ joints with $f_i$ DoF per joint using the formula $\#{\rm DoF}=3(n-1) -3 g +\sum_{i=1}^{g}f_i$. In Fig.~\ref{fig:fig1}$(b)$, by connecting two NSSs, we have $n=4$, $g=3$ and $f_i=1$ (corresponding to each pin joint), leading to a structure with three internal DoF. We visualize these DoF in Fig.~\ref{fig:fig1}$(b)$ as the shear, splay and compression modes of the planar NSS. These local modes of deformation of the planar unit cell when combined with the global assembly of NSS gives us the flexibility to create a new class of metamaterials. Although these structures can be deformed without an energetic cost they are not traditional mechanisms with floppy (zero energy) modes. Instead, as we will see, they have a different type of response close to that of ideal rigid-plastic solid, with one tunable parameter that is experimentally accessible.

As first examples of NSS assemblies, we show that we can design planar equipotential NSSs with varying Poisson's ratio (the negative ratio of the transverse strain  to the longitudinal strain) by simply varying the connectivity and orientation of the unit cells. In Fig.~\ref{fig:fig1}$(c)$ the structure has 1 DoF and a positive Poisson's ratio, in Fig.~\ref{fig:fig1}$(d)$ the structure has 8 DoF and zero Poisson's ratio, and in Fig.~\ref{fig:fig1}$(e)$ the structure has 5 DoF and a negative Poisson ratio. We note that the multiple internal DoF in Fig.~\ref{fig:fig1}$(d-e)$ are due the unconstrained internal modes of a subset of the assembly; later we will show how to arrest/control these. We can also assemble NS units into two-dimensional structures that have the ability to shift between different shapes (see SI Fig.~S2 and Fig.~S3).

To allow for the morphability of NSSs into three-dimensional structures starting from planar two-dimensional \textit{neutrally stable nets} (NS-nets), we replace the planar pin-joint between the lever and the strut in the unit-cell (see Fig.~\ref{fig:fig1}$(a)$) by a spherical joint, leading to a modified unit-cell shown in Fig.~\ref{fig:fig2}$(a)$ (see also SI Fig.~S4). Furthermore,  to ensure that the deformation of the NS-net only involves redistribution of the elastic energy stored in the prestretched elastic springs, we connect neighbouring struts (or levers) using NS joints (see SI Fig.~S7$(a)$) that allow for a single rotational DoF between two struts (or levers). Then, at each NS joint,   neighboring levers/struts can move in the local tangent-plane or normal to it. NS-nets thus represent a natural generalization of Tchebychev nets \cite{tchebychev1878coupe}  by including an additional local translational degree-of-freedom at the scale of the unit cell (see SI sec.~S6), in addition to the single orientational (shear) degree-of-freedom at every joint,. 

Figure~\ref{fig:fig2}$(a)$ shows an example of a patch of an NS-net consisting of six unit cells, each with a spherical joint between the lever and the strut; NS rotation at the spherical joints allows for local out-of-plane deformation of the patch. Just as in the planar case, the DoF of such a NS-net can be evaluated using the CGK formula \cite{marmier2016flexibility} that yields the expression DoF $=6(n-1) - 6g + \sum_{i=1}^{g}f_i$; thus, for the assembly shown in Fig.~\ref{fig:fig2}$(a)$, $n = 14$, $g = 15$, $f_i$ for each spherical joint is 3, and for each NS-pin joint is 1, yielding a total of 19 DoF.

To solve the inverse problem of designing NS structural materials that can morph from a planar periodic structure to a given complex 3D shape, we use an optimization approach similar to that recently deployed in other geometric optimization problems in origami and kirigami design \cite{van2017growth, dudte2016programming, choi2019programming}. Algorithmically, this corresponds to minimizing the shape mismatch error between the points on the target surface and equivalent points on the reference structures while satisfying the geometric constraints of neutral stability, implemented using the optimization routine \texttt{fmincon} in MATLAB. The cost function given an initial and target shape then takes the form

\begin{equation}
\E =\ \sum_{i,j} ||\aB^{(i,j)} - \tilde{\aB}^{(i,j)} ||^2 + ||\bB^{(i,j)} - \tilde{\bB}^{(i,j)} ||^2 \label{eq:2DEner}
\end{equation}
where $|| \A ||^2 = \tr^2 (\A) + \tr(\A^2)$, and  $\aB^{(i,j)}, \tilde{\aB}^{(i,j)}$ and  $\bB^{(i,j)}, \tilde{\bB}^{(i,j)}$ are the first and second fundamental forms~\cite{do2016differential} of the initial and target shapes evaluated at coordinate $\xB(u_j,v_j)$ (see SI sec.~S4 for details). Since all the deformations of NS-net are equipotential, one can realize multiple shapes with same net subjected to the local planarity constraints imposed by the NS pin-joints connecting the neighboring unit cells.

Figure~S4$(d)$ shows the ability of an NS-net to switch between two human faces with different geometric features. This is enabled by the capability of any patch of unit cells within an array to locally deform without affecting the neutral (mechanical) stability of the neighboring cells. The decoupling of the geometry from the mechanical response in NSS is in contrast with classical spring or elastomeric networks whose response to local perturbations is non-local and dimension dependent. We characterize the ability of an NS-net to morph into simple objects such as a cylinder, a sphere and a hyperboloid, and show the effect of varying the unit cell size on the shape morphing ability in SI Fig.~S4$(a-c)$.

\begin{figure*}[htb!]
\centering
\includegraphics[width=0.9\textwidth]{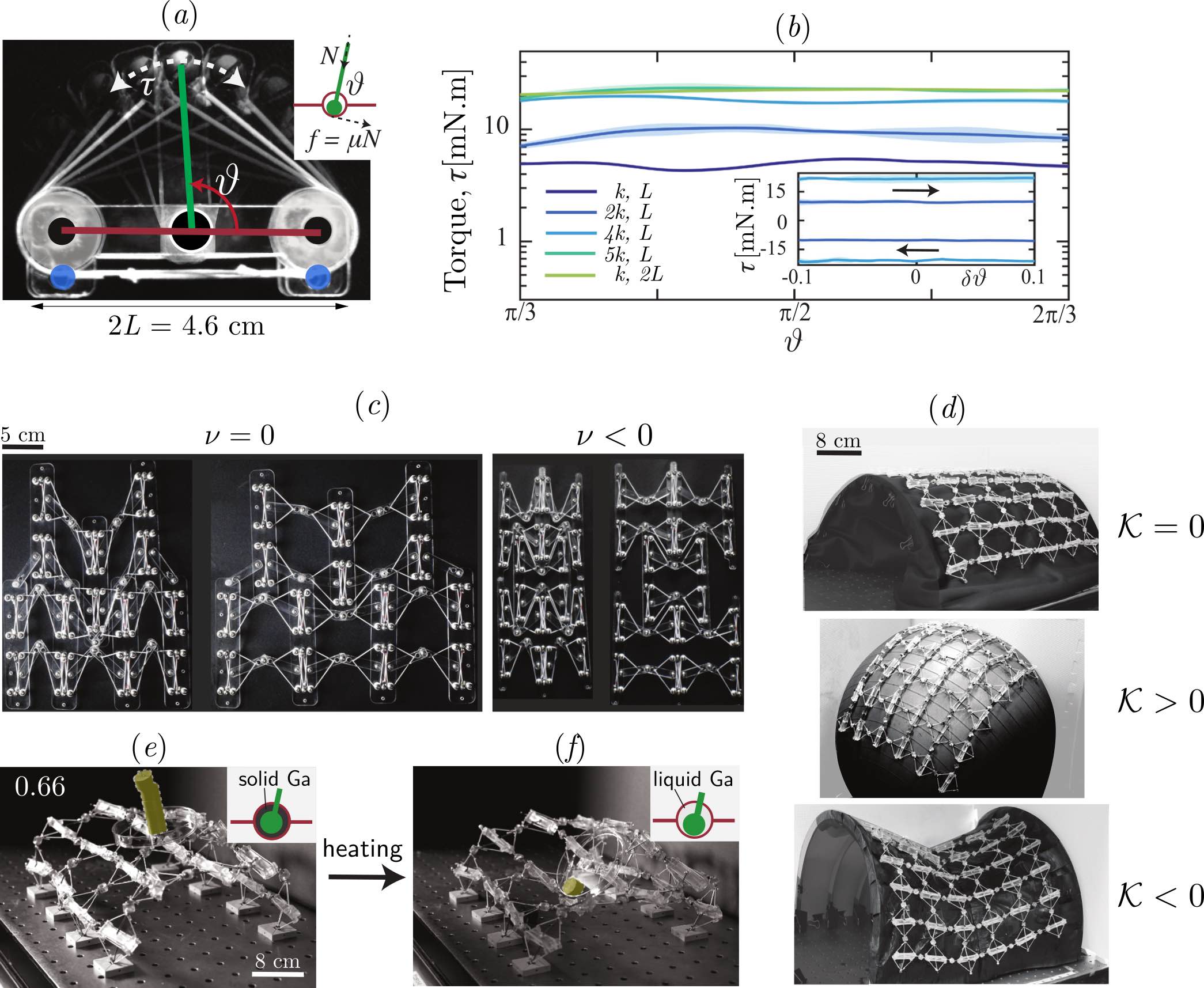}
\caption{\textbf{Experimental models of totimorphic neutrally-stable units and assemblies.} $(a)$ An experimental realization of a 2D neutrally stable unit cell with the lever and strut assembled using laser-cut plastic sheets. The composite image shows different equipotential configurations of the lever with respect to the base link. The black circle at the bottom of the lever indicates a pin joint and the blue circle indicates the location where the elastic strings are attached (see Fig.~S3 for further details). (inset) Schematic of the pin joint showing the internal frictional force that leads to a frictional torque. Although the configurations shown are equipotential, i.e. their elastic energies are equivalent, a finite  torque ($\tau$) is required to overcome internal frictional resistance.  $(b)$  The torque ($\tau$) as a function of angle $\vartheta$ for unit cells constructed with different spring constants $k$ and the geometric scale $L$, is nearly constant (shown on a log-linear scale). (inset) Constant torque is required to perturb the configuration  (near $\vartheta = \pi/2$). $(c)$ An experimentally realized assembly of 2D NSS with Poisson's ratio that is zero (left) and negative (right), just as theoretically predicted (Fig.~\ref{fig:fig1} $(d,e)$). $(d)$ Experimental realization of three-dimensional NS-net that morphs into surfaces with zero, positive and negative Gaussian curvature $\K$. $(e)$ A stimuli-responsive NS-net with Gallium-lubricated joints. When the Gallium is in its solid state, there is an increase in the internal friction in unit cells enabling a load carrying capacity (ratio of load carried to NS-net weight shown on top). $(f)$ Exposure to a higher temperature (35$^{\circ}$C) melts the Gallium in the joints, reducing the internal friction, and hence the threshold torque so that the structure collapses under the load.}
\label{fig:fig3}
\end{figure*}

 While NS-nets with periodic patterns of unit cells are capable of a range of shape-transforming capabilities, their ability to morph into shapes with large variations in curvature is limited by the size of the unit cell. To further the capability of NS-nets to morph into complex topographies with multi-scale features accurately, we need to construct non-uniform NS-nets with varying unit cell size. The problem of tessellating  the global minimization problem of solving for the entire 3D surface with NS cells in a single step (via energy minimization), is a non-convex (and elliptic) problem that arises from the constraints of neutral stability, and computationally expensive. An alternate approach of working with this constrained optimization problem is to convert the problem into a parabolic one using a marching method which tiles the surface in a step-wise manner, analogous to a recent additive approach to origami~\cite{dudte2021additive}.
The tiling process starts by discretizing a selected contour on the target surface into a set of edges of equal length (see SI Fig.~S5~$(a)$), each corresponding to the strut of a unit cell. We then construct the unit NS cells corresponding to these struts by finding the appropriate location of the levers. One end of the lever is at the midpoint of the strut and the other end is found by minimizing the vertex distance from the surface of the target shape while satisfying the constraints of neutral stability. This allows a selected contour on the surface of target object to become isomorphic to a connected layer of NS cells. We repeat this procedure with the optimized locations of the lever ends as the starting location to find the next connected layer and repeat the process until the entire surface is covered with NS units as shown in Fig.~S5~$(a)$. All the connections between individual unit cells are treated as planar NS joints whose plane of rotation is defined by the orientation of struts and levers on the tiled surface.

Figure~\ref{fig:fig2}~$(b)$ shows the NS-net tessellation for a catenoid, a surface with uniform negative Gauss curvature and zero mean curvature, using such a procedure. When the axis of tiling is along the direction of rapid change in the principal curvature, such as in the catenoid (see SI Fig.~S5$(b)$), our algorithm results in large number of units in the high curvature region. Beginning with a catenoid, a continuous deformation applied to all NS components morphs it to a helicoid with intermediate target shapes that are all minimal surfaces (ref Eq.~S7). Such a geometric deployment is possible without instabilities such as wrinkles on the NS-net because of the local in-plane shear DoF provided by the individual NS units as quantified in SI Fig.~S5~$(d)$. We note that the planar constraints between the neighbouring unit cells imposed by the NS pin joints restrict arbitrarily large deformations of the NS-net on the global scale and hence the final achievable shape of the helicoid has an expected roughness as evident in Fig.~\ref{fig:fig2}~$(b)$. Using the same algorithm we also tessellate an open cylinder, a surface with zero Gauss curvature and uniform mean curvature, and demonstrate its eversion in Fig.~\ref{fig:fig3}~$(d)$. We carry out this eversion by fixing the top of the cylinder while deforming the other end axially (details in SI sec.~S5). Note that all the partially everted states are energetically equivalent and stable. Similar to the catenoid transition, the local in-plane energy-free shear modes enables large deformations (shown in SI Fig.~S5~$(e)$).


To complement our computational results with physical realizations of NS-nets, we created neutrally stable unit cells using  laser-cut plastic sheets with the levers attached to the link through freely rotating pin joints, and elastic cords to mimic zero-length springs (see SI Fig.~S3).  NS unit cells/nets in our experiments differ from ideal floppy structures  as they are internally stressed. Furthermore, the finite size of the spherical and pin joints along with the presence of friction implies the need for a finite torque $\tau$ to rotate the struts at any joint by an angle $\vartheta$ in each unit cell (inset Fig.~\ref{fig:fig3}$(a)$). The value of this torque can be controlled by changing either the spring stiffness or the lever length within a unit cell, since this changes the frictional forces at the joints and thence the torque.   In Fig.~\ref{fig:fig3}$(b)$ we see a proportional increase in the torque with increasing spring stiffness $k$ and lever length $L$; furthermore we see that the torque is relatively independent of $\vartheta$ over a range of angles
(see Fig.~\ref{fig:fig3}$(b)$ inset) (see also SI Fig.~S8-S9). This response is reminiscent of a rigid-plastic material~\cite{hill1998mathematical}, for long just a mathematical idealization of plastic behavior in solids, but one that arises here physically as a consequence of the internal friction at the joints.

In Fig.~\ref{fig:fig3}$(c)$ we show physical realizations of planar NS-nets with both zero and negative Poisson's ratio that follow their theoretical idealizations in Fig.~\ref{fig:fig1}$(d,e)$ (see SI Fig.~S3$(d, e)$ for details of their mechanical response). To show how to achieve a shape morphing NS-net in three dimensions, we use ball and socket joints to connect the levers and struts within a unit cell and also to connect neighboring unit cells (see SI Fig.~S3$(f)$). This allows us to fabricate NS-nets that can form surfaces of zero Gauss curvature, e.g. a cylinder, positive Gauss curvature, e.g. a sphere and negative Gauss curvature, e.g. a hyperboloid as shown in Fig.~\ref{fig:fig3}$(d)$. In all these cases, the deformations of the NS-net are primarily in the joints connecting nearest neighbor NS units, just as shown in Fig.~\ref{fig:fig2}$(b)$.

Although friction in the NS-net enables a rigid-plastic response, the structural materials made from NS unit cells do not have the ability to support substantial loads owing to the large number of internal degrees of freedom. To overcome this limitation and to move towards a NS-net which can be rigidified on-demand while retaining the flexibility that arises from neutral stability, we use the phase-changing property of gallium at near room temperatures (melting point of $30^{\circ}$C)  and fill all the ball joints with gallium~\cite{ye2016phase}.  Gallium can then act either as a viscous lubricant in the ball joints in its liquid state (at room temperature), or as a solid that impedes rotation (at low temperatures). Thus by controlling the local joint temperature, we can make the NS-net multimorphable when the gallium is in liquid state, or stiff when the gallium is in its solid state. This ability to rigidify the NS-net allows it to support external loads as shown in Fig.~\ref{fig:fig3}$(e)$ where a cylindrical NS-net can carry a load until it is heated to about ~$\sim 35^{\circ}$C when the structure buckles and collapses (Fig.~\ref{fig:fig3}$(f)$).

Our approach of using neutrally stable unit cell based assemblies offers a simple approach to totimorphic assemblies by separating the geometry of the assembly from its mechanical response at both the individual and collective level. The local geometry of the unit cell can be  varied by changing both its overall size as well as the length of the single movable strut, while its plastic response can be changed by varying either the stiffness of the springs within the structure or the length of the struts and links. This allows for individual neutrally stable structures to be built on any scale, and then assembled into structures by modulating the size and stress within a unit cell to create spatially heterogeneous material structures. At a practical level, the mechanical response of individual cells can be further controlled by tuning the rheological response of the pin/spherical joints by using a phase-changing material such as gallium. All together, this allows for totimorphic NS assemblies to simultaneously have both local structural flexibility as well as a heterogeneous mechanical response.

\begin{acknowledgments}
{We thank the Whitesides and Bertoldi Labs at Harvard University for sharing experimental resources, and Amit Nagarkar for help with the Gallium joints. The work was supported partially by NSF grants BioMatter DMR 1922321 and MRSEC DMR 2011754 and EFRI 1830901.}
\end{acknowledgments}

\bibliographystyle{unsrtnat}
\bibliography{pnas-sample}

	\large
	\widetext
	\clearpage
\onecolumngrid
\begin{center}
\textbf{\large Supplemental Materials: Totimorphic assemblies from neutrally-stable units}\\[.2cm]
Gaurav Chaudhary,$^{1, *}$ S  Ganga  Prasath,$^{1, *}$ Edward Soucy$^{2}$ and  L  Mahadevan$^{1, 2, 3}$\\[.1cm]
{\small \itshape ${}^1$School of Engineering and Applied Sciences, Harvard University, Cambridge MA 02138.\\
${}^2$Center for Brain Science, Harvard University, Cambridge, MA 02138.\\
${}^3$Department of Physics, Harvard University, Cambridge MA 02138.\\
${}^4$Department of Organismic and Evolutionary Biology, Harvard University, Cambridge 02138.\\
${}^*$equal contribution}
\end{center}
	\setcounter{equation}{0}
	\setcounter{figure}{0}
	\setcounter{table}{0}
	\setcounter{page}{1}
	\makeatletter
	\renewcommand{\theequation}{S\arabic{equation}}
	\renewcommand{\thefigure}{S\arabic{figure}}
	\renewcommand{\bibnumfmt}[1]{[#1]}
	\renewcommand{\citenumfont}[1]{#1}
	\linespread{1.5}

	\section{Planar equipotential transformations}
\label{sec:poisson}
\begin{figure}[h!]
\centering
\includegraphics[width=\textwidth]{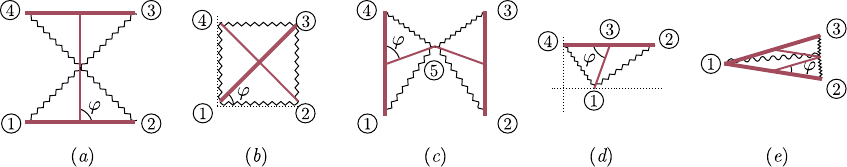}
\caption{Sub-units of assembly shown in Fig.~1 of main text used to calculate the transformation matrix for $(a)$ simple shear, $(b)$ positive Poisson structure, $(c)$ zero Poisson structure, $(d)$ auxetic structure, and $(e)$ chiral and achiral structure. Each unit shown here can be assembled to generate the macroscopic assembly to get peculiar shear, Poisson's ratio and chirality.}
\label{fig:suppl1}
\end{figure}
Here we calculate the transformation matrix which defines the motion of assembly under Neutrally Stable constraints and the Poisson's ratio of the material. In order to do this we consider minimal units from the the assembly shown in Fig.~1 of the main text whose in-plane affine transformations is enough to describe the entire material behaviour. Let us define the transformation operators useful to describe the motion of 2D NS metamaterials. The translation matrix acts on a position vector to translate it by a supplied vector: $\T(\xB; \v) \equiv \T(\v) = [1\ 0\ v_x; 0\ 1\ v_y; 0\ 0\ 1] \cdot \xB = \xB + \v$, while the rotation matrix results rotation by an angle: $\Rbd(\xB; \phi) = [\cos \phi -\sin \phi; \sin \phi \cos \phi] \cdot \xB $. The global motion of each assembly is captured using these in the following manner.

\begin{figure}[h!]
\centering
\includegraphics[width=\textwidth]{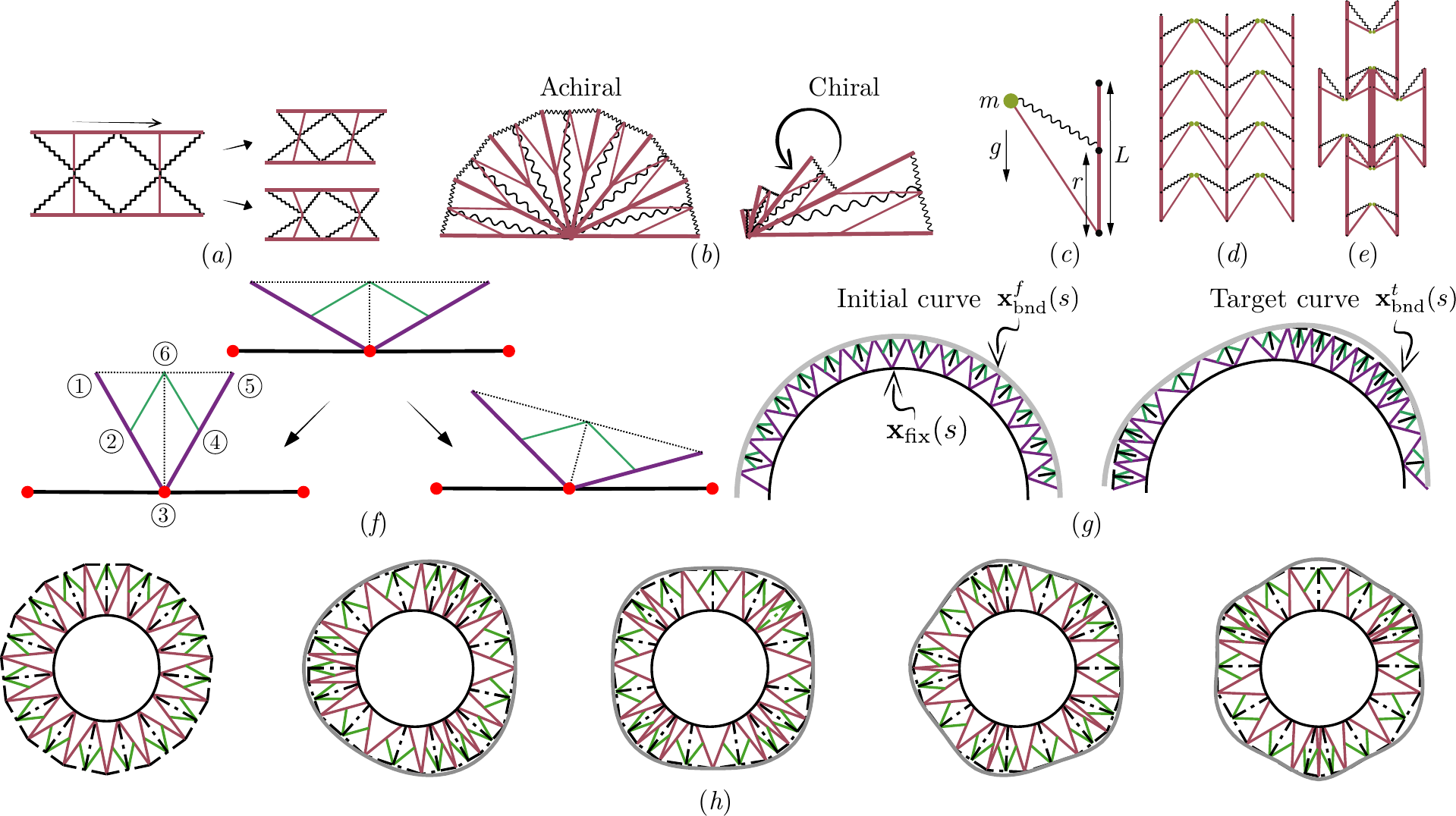}
\caption{$(a)$ Assembly of NS units showing extreme configurations of the levers when acted under shear. $(b)$ Achiral material with axis of symmetry and material with intrinsic chirality. $(c)$ A single unit of gravity equilibrated structure with mass $m$ attached to the end of a lever of length $L$, assembled to a $(d)$ zero Possion structure and $(e)$ negative Poisson structure. Schematic showing $(f)$ the two modes of NS deformation that the transformation exploits to shift between curves, $(g)$ the setup for transforming boundary in 2D. $\xB_\fix(s)$ - fixed inner boundary around which the one end of the NS structure slips; $\xB^{f}_\bnd(s)$ - initial boundary, $\xB^{t}_\bnd(s)$ - target shape to which the assemble transforms. $(h)$ Equipotential shape transformations of a 2D NS assembled into an annulus. The inner boundary is held fixed with the unit cells allowed to slip. The 2D assembly can transform between various initial and target shapes.}
\label{fig:suppl2}
\end{figure}

\begin{itemize}
\item \textbf{Simple shear}: The initial coordinates of the unit shown in Fig.~\ref{fig:suppl1}$(a)$ is $\{\xB_1, \xB_2, \xB_3, \xB_4 \} = \{ (0,0), (2l, 0), (2l, 2l), (0, 2l) \}$. This unit is a minimal representation enough to capture deformations of the assembly shown in Fig.~\ref{fig:suppl2}$(a)$. Under shear transformation, the states are given by change in angle $\varphi$. This leads to coordinates as a function of angle $\varphi$ given by $\P(\varphi) = \{ \xB_1, \xB_2, \T(\xB_3; (2l \cos \varphi, 2l(1 - \sin \varphi)), \T(\xB_4; (2l \cos \varphi, 2l(1 - \sin \varphi))) \}$.
\item \textbf{Positive Poisson structure}: Similar to the simple shear case, we can now write the transformation matrix the structure in Fig.~1$(c)$ for a given angle $\varphi$ (ref Fig.~\ref{fig:suppl1}$(b)$) as $\{\xB_1, \xB_2, \xB_3, \xB_4 \}$ = $\{ (0,0),$ $(\sqrt{2}l, 0),$ $(\sqrt{2}l, \sqrt{2}l),$ $(0, 2l) \}$ as $\P(\varphi) = \{ \xB_1,$ $\T(\xB_2; (2 l \cos \varphi, 0)),$ $\Rbd(\xB_3; \varphi),$ $\T(\xB_4; (0, 2l \cos \varphi)) \}$.

We further define the Poisson's ratio for the 2D structure in Fig.~1$(c)$ as a function of $\varphi$. We have:
\[
\nu = -\frac{\varepsilon_\trans}{\varepsilon_\axial}
\]
where
\begin{align*}
\varepsilon_\axial =& \ \frac{\delta l_\x}{l_\x}, l_\x =\ 4L \cos \phi, \delta l_\x =\ l_\x(\varphi + \delta \varphi) - l_\x(\varphi) = \ -4L \sin(\phi) \delta \phi.\\
\varepsilon_\trans =& \ \frac{\delta l_\y}{l_\y}, l_\y =\ 4L \sin \phi, \\
\delta l_\y =& \ l_\y(\varphi + \delta \varphi) - l_\y(\varphi) = \ 4L \cos(\phi) \delta \phi.\\
\implies \nu =& \ \cot^2 \phi.
\end{align*}

\item \textbf{Zero Poisson structure}: The initial configuration (ref Fig.~\ref{fig:suppl1}$(c)$) that minimally represents the assembly in Fig.~1$(d)$  is given by $\{\xB_1, \xB_2, \xB_3, \xB_4, \xB_5 \} = \{(0,0), (2l, 0),$ $(2l, 2l), (0, 2l),$ $(l, l) \}$. The transformation of the structure can then be written as: $\P(\varphi) = \{ \xB_1, \T(\xB_2; (0, 2 l (1 - \sin \varphi)), \T(\xB_3; (0, 2 l (1 - \sin \varphi)), \xB_4, \T(\xB_5; (0, l(1 - \cos \varphi)) \}$.

\item \textbf{Auxetic structure}:
The transformation of the auxetic assembly in Fig.~1$(e)$ from initial position (ref Fig.~\ref{fig:suppl1}$(d)$) $\{\xB_1, \xB_2, \xB_3, \xB_4 \} = \{(l,0), (2l, l), (l, l), (0, l)\}$ is given by: $\P(\varphi) = \{ \T(\xB_1; (l(1-\cos \theta),0))$, $\T(\xB_2; (0, l \sin \varphi))$, $\T(\xB_3; (0, l \sin \varphi))$, $\T(\xB_4; (0, l \sin \varphi)) \}$
Similarly, we have for the auxetic structure:
\begin{align}
\varepsilon_\axial =& \ \frac{\delta l_\x}{l_\x}, \\
l_\x =& \ 2L (1- \cos \phi),	\delta l_\x =\ l_\x(\varphi + \delta \varphi) - l_\x(\varphi) =\ 2L \sin \varphi \delta \varphi. \\
\varepsilon_\trans =& \ \frac{\delta l_\y}{l_\y},\\
l_\y =& \ 2L \sin \varphi, \delta l_\y =\ l_\y(\varphi + \delta \varphi) - l_\y(\varphi)	=\ 2L \cos(\phi) \delta \phi.\\
\implies \nu =& \ -\frac{\cot \phi}{\cot (\phi/2)}.
\end{align}

\item \textbf{Achiral and chiral structure}:
For the achiral structure (see Fig.~\ref{fig:suppl2}$(b)$) we can write the transformation from the initial position (ref Fig.~\ref{fig:suppl1}$(e)$) given by: $\xB_j = 2l$ for $j=2, 3$ and $\xB_1 = (0, 0)$ as $\P(\varphi) = \{ (0,0), \Rbd (\xB_2; -\varphi/2), \Rbd (\xB_3; \varphi/2) \} $. The transformation remains the same for the chiral structure, however now the length varies for each sub-unit.
\end{itemize}

\section{Gravity equilibrator}
The classical Anglepoise lamp is an example of a gravity equilibrated structure where the geometry of the structure is tuned such that all deformations of the mass keep the total energy of the system a constant. This is achieved by exchange of energy from potential energy of the attached end mass and the stretching energy of the point spring~\cite{herder2001energy,carwardine1932improvements}. The simplest such assembly is shown in Fig.~\ref{fig:suppl2}$(c)$ where the mass $m$ is attached by a point spring of spring constant $k$ through a lever of length $r$. The deformation of the mass becomes independent of the orientation when the mass satisfies: $m g = kr$. This unit structure can then be assembled into either a zero or a negative Poisson material by connecting into structures shown in Fig.~\ref{fig:suppl2}$(d, e)$. It is worth mentioning that though the unit cells themselves contribute a total weight from the top layer that get added to the layers below them, this effect can be overcome if the entire structure is mounted on a stage such that the weight of the structure doesn't affect others below them.

\section{Numerical methods for shape morphing in 2D and 3D} \label{sec:numMeth}
\subsection{In 2D}
\label{sec:2Dcurve}

\begin{figure*}[h!]
\centering
\includegraphics[width=\textwidth]{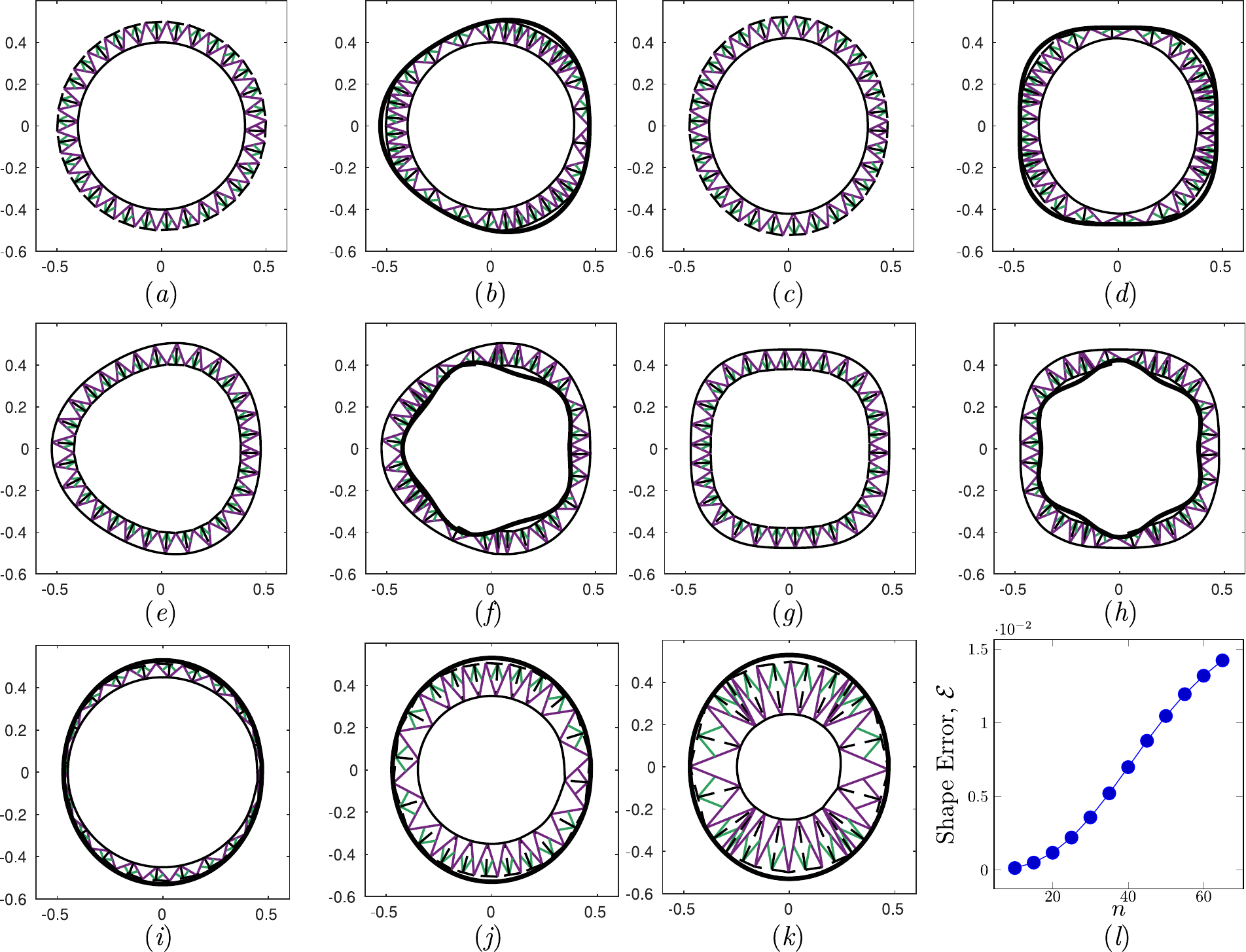}
\caption{$(a, c, e, g)$ Initial for different values of $m_f = 1, 2, 3, 4$ and $(b, d, f, h)$ target shapes for different values of $m_t = 3, 4, 5, 6$ for a fixed number of NS units, $n = 30$. We show that the target shapes can be such that the assembly can transform both the inner and outer boundary. $(i, j, k)$ We show target shapes for $r_f = 0.45, 0.35, 0.25$ while fixing the outer radius as $r_t = 0.5$, for fixed $n = 30$. $(l)$ Error in achieving the target shape as a function of number of NS units, $n$.}
\label{fig:suppl3}
\end{figure*}

Now that we know the constraints of each NS unit, we move towards the algorithm to shift between two different curves (initial curve and target curve), which is performed in two steps. First step is to create an assembly which matches exactly the shape of the initial curve parameterized as $\xB^{f}_\bnd(s)$ and then we deform this curve to a target curve $\xB^{t}_\bnd(s)$ approximately. The assembly itself is composed of two ends, one end confined to a curve $\xB_\fix(s)$ that can slip freely along this curve and the other end which matches with $\xB^{f}_\bnd(s)$ and $\xB^{t}_\bnd(s)$ (see Fig.~\ref{fig:suppl2}$(g)$). The fixed curve is parameterized using $\xB_\fix(s)$ where the base of the NS structure, $\xB_i^{(3)}$ (see Fig.~\ref{fig:suppl2}$(f)$ for the indices) lies and the location of its ends, $\xB^{(1),(5),(6)}$ must lie along $\xB^{f}_\bnd(s)$. In order to do find such a shape, we use $l$, the length of the lever or the half length of the link as a free variable and solve the constraint problem of $\xB_i^{(3)}$ lying along $\xB_\fix(s_j)$ and $\xB_i^{(1),(5),(6)}$ along $\xB^{f}_\bnd(s_j)$. From this initial shape we transform to $\xB^{t}_\bnd(s_j)$ approximately by minimising an energy, which is equivalently a measure of the shape error, defined as:
\[
\E = \sum_{i} \sum_{j=1, 5, 6} || \xB^{t}_\bnd(s_j) - \xB_i^{(j)} || \ s.t. \quad g_{NS}(\xB_i,l) = 0.
\]
Thus there is an error associated with the transformation from the first curve to the second one. The error in shape shifting is primarily a resultant of the non-intersecting condition on all the elements in unit cells. As a demonstration of the shape-transforming capability, we show how to construct annuli with varying azimuthal wave numbers.  In Fig.~\ref{fig:suppl2}$(h)$ we show the fixed curve $\xB_\fix(s)$ in solid black and the curve the target curve $\xB^{t}_\bnd(s)$ is shown in gray. Figure~1$(f)$ are the initial shapes $\xB^{f}_\bnd(s), \xB_\fix(s)$ given by circles which transform to target shapes of the form $\xB^t_\bnd(s) = r_t(1\pm\cos (m_t s))(\cos \theta(s), \sin \theta(s))$ with $m_t=3$ in $(b,e)$, $m_t=5$ in $(c,f)$. We also consider $\xB_\fix(s) =  r_f(1\pm\cos (m_f s))(\cos \theta(s), \sin \theta(s))$ and in Fig.~\ref{fig:suppl3}$(a-k)$ plot solutions obtained for different combinations of $m_f, m_t$ and $r_f, r_t$. Energy $\E$ is an estimate of the shape error between the target shape and the approximately transformed shape. This is shown as a function of number of NS units in Fig.~\ref{fig:suppl3}$(l)$.


\subsection{In 3D}
In order to match the shape of the surface to the target shape, we minimize the energy described in Eq.~1. Each neutrally stable unit is represented by positions $\xB^{(i)}, i = 1, 2, 3, 4$ shown in Fig.~1$(a)$. They have to satisfy rigid constraints $g_{NS}(\xB,l) \equiv g(1,2;l) = g(2,3;l) = g(2,4;l) = g(1,3;2l) = 0$, where we define:
\begin{align*}
g(i,j;l) \equiv ||\xB^{(i)}-\xB^{(j)}||^2 - l^2 =& \ 0.
\end{align*}
Here $l$ is the length of the lever and the link is twice this length. Each unit cell satisfying this constraint has its rotation for free.
\blue{In 3D the complexities arising out of the third axis is considerably different from that of 2D. In 2D in Fig.~1$(b)$ we saw that two unit cells connected by a pin-joint can have 3 modes of motion. However in 3D, the various combinations arising out of different possibilities of joints makes the problem complex. We extend the 2D framework to 3D where we make the following assumptions:}
\begin{itemize}
    \item \blue{Connection between two levers and two struts have only one allowed angle of rotation. Let $\th_1$ be the orientation of one of the levers/struts and $\th_2$ be the orientation of the other lever/strut, then they are allowed to rotate either around $\bh = \th_1 \times \th_2$ or $\dh = \th_1 \times \bh$ (see Fig.~2~$(a)$).}

    \blue{This constraint of rotational DoF implies that there exists a NS-joint that allows such a transformation and prevents the structure from being floppy. In order to connect two neighboring NS units via struts or via levers, we introduce a joint that is neutrally stable and can distribute angular deformations of the connecting links to the strain energy of the elastic strings. This is done through an assembly shown in Fig.~\ref{fig:suppl5}$(a)$ where we have two mini-links with a lever each, connected by a pin-joint that couples the angular deformation of the mini-link with two elastic strings attached to each of the mini-links. As the relative angle of the mini-link changes, the elastic energy stored in both the strings in each of the mini-links. We can connect this joint along the axis of the lever/strut or perpendicular to it thus allowing rotation along either $\bh$ or $\dh$.}
    \item \blue{All the levers are connected to the strut via a spherical joint with 3 DoF i.e. rotation along three axis of the lever.}
\end{itemize}
\blue{Using these assumptions, each lever/strut joint has two combinations and thus a $2 \times 2$ grid of NS unit cells composed of 2 pairs of struts, 2 pairs of levers have a total of 12 unique combinations. However we restrict our analysis here to a specific set of combinations that is sufficient to create 3D nets with gaussian curvature $\K >/=/< 0$.}

\subsubsection{Transformation of a planar NS-net to cylinder, sphere and saddle}
\label{sec:planeSphere}
In the 3D scenario, we first tile a plane with NS unit cells in a grid as the initial shape. To keep track of all the components we convert the whole structure into a triangulation data structure in MATLAB.
To transform the planar structure into a 3D surface, we first find a target shape by mapping each node. The shapes described in Fig.~2 are of the form $\vec{f}(u,v)$. For all these three shapes we evaluate the first and second fundamental forms given by $\aBt^{(k,l)}, \bBt^{(k,l)}$ at each node $\xB(x_k,y_l)$ where $k, l$ are the indices for $x$-location and $y$-location. Here the target fundamental forms $\aBt = ( E~F; F~G )$ and $\bBt = ( L~M; M~N )$ are with the following functional forms:
\begin{itemize}
\item Cylinder:
\begin{align*}
\vec{f}(u,v) =&  \{ \cos{u}, \sin{u}, v\} \\
\{E, F, G\} =&  \{ 1, 0, 1\} \\
\{L, M, N\} = & \{ 0, 0, 1\} \\
\text{for } v \in [0, \pi] & \text{ and } u \in [0, \pi]
\end{align*}

\item Sphere:
\begin{align*}
\vec{f}=& \{\sqrt{1-u^2}\cos{v}, \sqrt{1-u^2}\sin{v}, u \} \\
\{E, F, G \} =&  \{ \sin^2{v}, 0, 1\} \\
\{L, M, N\} = & \{ \sin^2{v}, 0, 1\} \\
\text{for } v \in [-1, 1] & \text{ and } u \in [-1, 1]
\end{align*}

\item Hyperbolic paraboloid (saddle)
\begin{align*}
\vec{f} =& \{u, v, u^2 - v^2 \} \\
\{E, F, G \} =& \ \{ 1 + v^2, u v, 1 + u^2\} \\
\{L, M, N\} = & \frac{1}{\sqrt{1+u^2 +v^2}}\{ 0, 1,  0\} \\
\text{for } v \in [-1, 1] & \text{ and } u \in [-1, 1]
\end{align*}
\end{itemize}

For each case, we minimize the error in the transformed shape by minimizing the shape error given by:
\begin{align}
\E =& \ \sum_{i,j} ||\aB^{(i,j)} - \tilde{\aB}^{(i,j)} ||^2 + ||\bB^{(i,j)} - \tilde{\bB}^{(i,j)} ||^2
\end{align}
which is the L2-norm on invariants of the difference in fundamental forms where $|| \A ||^2 = \tr ^2 (\A) + \tr(\A^2)$. This is evaluated for each cell $\xB^{(j)}(x_k,y_l)$ for intermediate shapes while simultaneously satisfying the NS constraints $g_{NS}(\xB^{(j)}(x_k,y_l)) = 0$. The cases shown in the main article were solved on a Intel i9 computer with a RAM of 32 GB using MATLAB's constrained optimisation routine \texttt{fmincon}. The geometric constraints for maintaining the neutral stability of the structure is supplied as a constraint to the minimization problem in $\texttt{fmincon}$. In Fig.~\ref{fig:suppl6}$(a-c)$ such a transformations are performed, and the results are shown.

\subsubsection{Shape morphing between Maxwell and Gauss face}\label{sec:faceMorphing} 
\begin{figure*}[h!]
\centering
\includegraphics[width=0.75\textwidth]{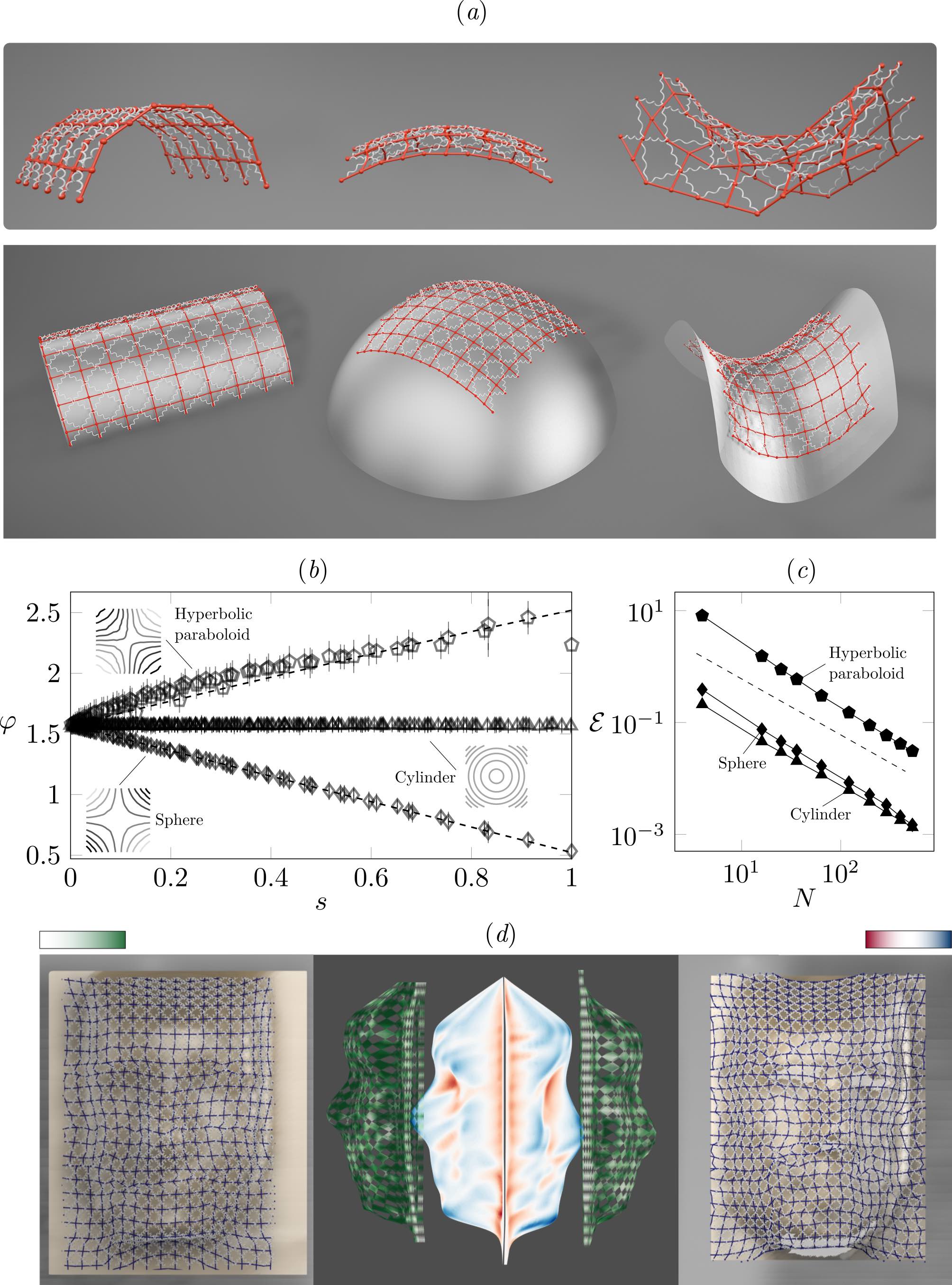}
\caption{$(a)$ A planar array of NS structures morphs into three different shapes with intrinsic curvature: a cylinder ($\K=0$), part of a sphere ($\K > 0$), and hyperbolic paraboloid ($\K < 0$). The top simulations correspond to NS joints without any approximation i.e. with only one degree of freedom and the bottom one with the spherical joint approximation. $(b)$ Angle $\phi$ made by the NS assembly as a function of $s = uv$ where $u$ is the parameterization along the rigid link and $v$ along ends connecting two NS structures (shown in Fig.~1$(b)$). Comparison with the solution for Tchebychev nets for which both the surface preserves length both $u$ and $v$ is shown with dashed lines. $(c)$ The root mean square error between the morphed shape and the target decays to zero with increasing number of unit cells. $(d)$ \blue{Shapes with complex mean curvature (shown in the middle) such as the faces of James Clerk Maxwell (left) and Carl Friedrich Gauss (right) can be realized from the same planar NS array. The neutral stability of NS array in all configurations allows morphing into such complex shapes, as shown in the extreme left and right images. The NS array has  $60 \times 60$ NS units. Error in shape morphing defined as the absolute error in mean curvature at every point of the array normalized by the root mean squared val ue of target curvature is shown on the colormaps next to the respective target shapes.}}
\label{fig:suppl6}
\end{figure*}

In order to extract the face of the physicist J.C. Maxwell and the mathematician C.F. Gauss, we use the code developed in \cite{jackson2017large} from which we extract the 3d mesh with from a 2d picture. We then interpolate the coordinate along a uniform grid, and scale down the coordinates between [0,1]. This grid is then used as a target shape. Using the same technique as above, we perform this transformation by minimising the same shape error $\E$. In Fig.~\ref{fig:suppl6}$(a)$ such a transformation is performed for $n=50\times50$.

\section{Deployment strategies for morphing NS-net}\label{sec:deployment}

\begin{figure*}[h!]
\centering
\includegraphics[width=0.9\textwidth]{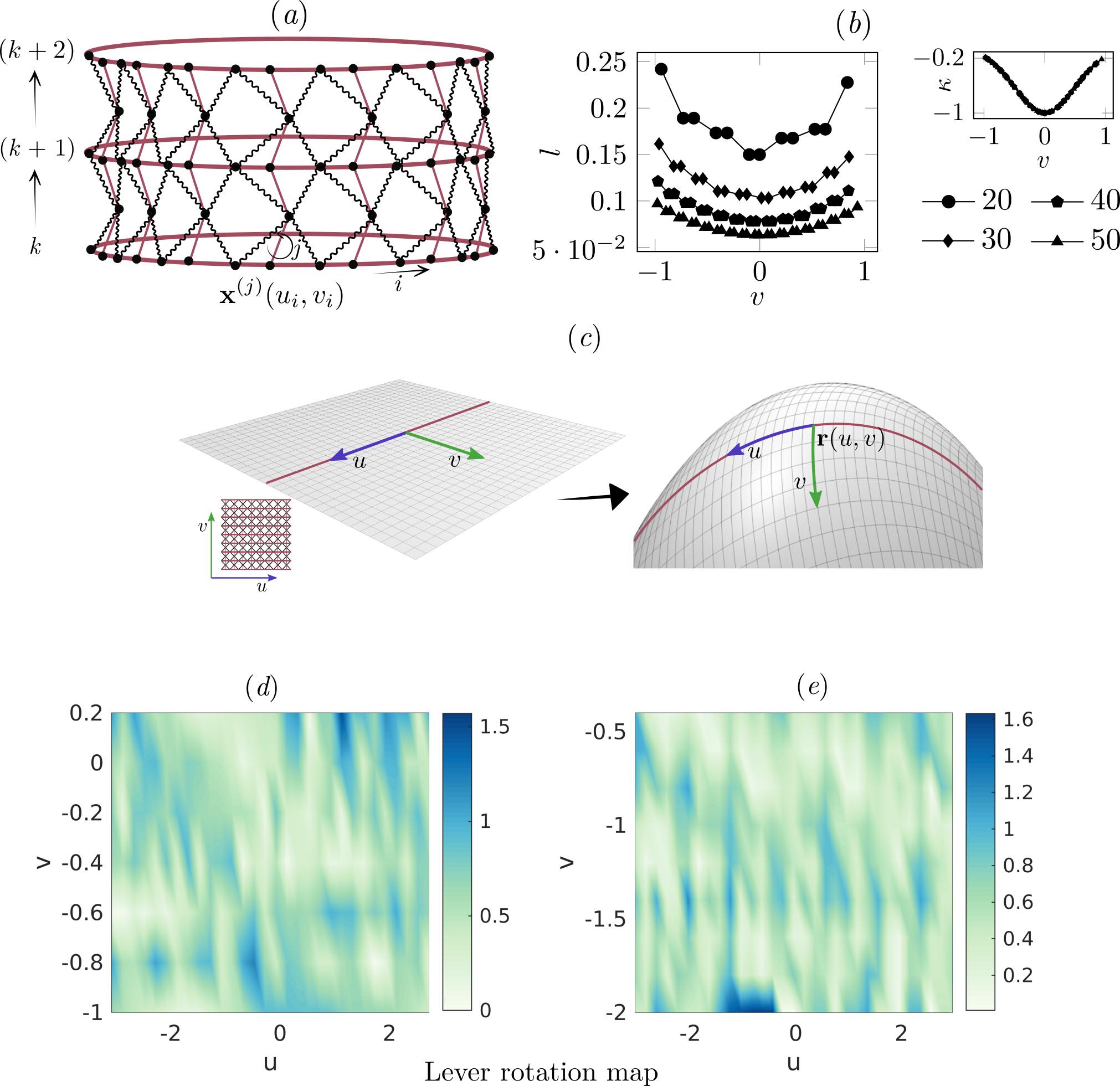}
\caption{$(a)$ Step-wise tessellations of an object is performed by discretizing a chosen curve $(i)$ on the surface with links of unit cells, and marching along a chosen direction ($k \to k+1$) to find the location of lever ends that satisfy the geometrical constraints of NS, and lie on object's surface. In the following step, the link ends for the next layer ($k+1$) are found with same set of constraints. $(b)$ The length of lever $(l)$ along a vertical section of a catenoid tessellation (shown in (c)) for different number of discretizations of the initially chosen curve (base of catenoid ($v=-1$). The high curvature region of the catenoid is naturally discretized with a finer resolution (l). $(c)$ Mapping of a plane in $\R^3$ parameterized by $u, v$, where the inextensible lines are along $u = \text{const.}$ and this is morphed to a surface with coordinates $\r(u,v)$. $(d-e)$ The deformation of a catenoid to a helicoid utilizes the NS lever rotation w.r.t to the strut. Here we plot the lever rotation map quantified by the change in the orientation of the lever w.r.t. the struct in the NS-net for $(d)$ a saddle to helocoid transition, and (e) a cylinder eversion. The $u-v$ axis are the parameterizations that map a plane to a catenoid and sphere respectively. }
\label{fig:suppl7}
\end{figure*}

\subsection{Catenoid to helicoid}\label{ssec:heltocat} In order to tile the surface of a catenoid, we use a stepwise local tiling strategy instead of formulating the problem as a global constraint problem, for which finding a solution is difficult as it is NP-hard. What we mean by this is that, since we would like to fill a given surface using NS units, which as we have shown boils down to finding solution to the rigid constraints with a defined connectivity between NS units.  Aside from this we further have the non-overlapping constraints. This if we would like to do for arbitrary surfaces, the number of non-linear constraints along go as $\O(5n \times N)$ where $n$ is the number of cells in a layer and $N$ is the total number of number layers. In order to avoid this trap, we hover by formulating the programming problem as a local problem. Under this strategy we start with one of the boundaries (also feasible to start from the interior of the surface) denoted by layer $k$ whose coordinates are $\xB^{(j)}_{(k)}(u_i,v_i)$. Here $j = 1, 2, 3, 4$ denotes the coordinates associated with boundary coordinates of the $i$-th cell in the $k$-th layer. Each $k$-th layer (see Fig.~\ref{fig:suppl7}$(a)$) has $n$ cells thus $i = 1, \dots, n$ and the layers go between $k = 1, \dots, N$. Now we need to find coordinates $\xB^{(j)}_{(k+1)}(u_i,v_i)$ in the $(k+1)$-th layer which satisfy the condition $\xB^{(4)}_{(k)}(u_i,v_i)=\xB^{(4)}_{(k+1)}(u_i,v_i)$. We have however not fixed $l$ for each $i$, in both the $k, (k+1)$ layers which is found by solving the local constraint problem. By writing it as a local problem we have converted number of constraints from $\O(5n \times N)$ to $N \times\O(5n)$, which is far simpler to solve. We show the efficacy of our algorithm by transforming from a catenoid to a helicoid which has a classical analytical solution given by $\xB(u,v) = (x(u,v), y(u,v), z(u,v))$ where:
\begin{align}
x(u,v) =& \ \sin t \,\cosh v\,\cos u + \cos t \,\sinh v\,\sin u, \label{eq:hel2cat}\\
y(u,v) =& \ \sin t \,\cosh v\,\sin u - \cos t \,\sinh v\,\cos u, \\
z(u,v) =& \ u\cos t +v\sin t.
\end{align}
such that $(u,v) \in (-\pi ,\pi ] \times [-L, L]$ and the deformation parameter $-\pi <t \leq \pi $. In Fig.~2$(b)$ we have used to $n = 47, N = 8$.

\subsection{Eversion of a cylinder}\label{sec:cylconsph}
In the previous example we showed that we can transform surfaces using their parameterization. We can also achieve similar motion by prescribing boundary displacement. We can consider an examples of inverting a cylinder (see Fig.~2$(d)$ of the main text). We first tile the cylinder with NS structures using the same technique detailed in the previous subsection. Now we fix one end(top) of the cylinder and radially shrink the bottom boundary (alternate NS pin-joints between the levers of the bottom most layer) from an initial radius of $a_f=0.4$ to $a_f=0.3$. Then we move this boundary along a vertical trajectory in small steps (20) until the cylinder is turned inside out.

\section{Geometry of NS materials}
\label{sec:gNS}
Earlier we developed a framework to program NS structures such that they achieve a target shape using discrete collection of links and levers (connected by springs). When the number of NS units are large, we can represent NS structures as a continuous surface in $\R^3$ where the lines representing the parameterization along one direction become the inextensible links of the NS structure while the other direction becomes the region with springs. Let us consider a plane in $\R^3$ parameterized by $u, v$ where $u = \const$ and $v = \const$ are perpendicular to each other~\cite{do2016differential}. For NS structures, we assume that the inextensible structures are located along $v = \const$ (see Fig.~\ref{fig:suppl6}$(d)$). When we deform this planar configuration to a target shape $\r(u,v)$, NS structures leverage the two modes of deformation already described, i.e. compression and shear. In the ensuing section we show that in order to find the shape of transformed NS materials, all we need to do is represent the target surface using a parameterization such that along $v = \const$ curves preserve length. It is worth mentioning here that since the number of unit cells is infinity, the number of free modes is also infinity in the continuum setting. Thus it should come as no surprise that we can wrap an NS material on to any surface.

\subsection{Surface representation}
Let the target surface $\Sigma$ in $\R^3$ parameterised by $\r(u,v)$ have a unit normal $\N(u,v)$ given by
\[
\N = \frac{\r_u \times \r_v}{| \r_u \times \r_v |}.
\]
Though $u, v$ are as yet independent parameterization, later we will see that they are nothing but the parameterization of the initial planar configuration. The surface $\Sigma$ has the first and the second fundamental form as
\begin{align}
I =& \ \d \r \cdot \d \r = E\ \d u^2 + 2 F \ \d u \d v + G\ \d v^2,\\
II =& \ -\d \r \cdot \d \N = e\ \d u^2 + 2 f \ \d u \d v + g\ \d v^2.
\end{align}
Here $E = \r_u \cdot \r_u, F = \r_u \cdot \r_v, G = \r_v \cdot \r_v, e = -\r_u \cdot \N_u, f = -\r_u \cdot \N_v, g = -\r_v \cdot \N_v$. The sextuplet $\{ E, F, G, e, f, g \}$ describes the surface $\Sigma$ up to position in space. Since we are interested in NS structures described using lines, we define $\a = \r_u$ the tangent vector along $v = \text{const.}$ and $\b = \r_v$ along $u = \text{const.}$, such that for the network we have: $d\r = \a \ \d u + \b \ \d v$. These vectors also need to satisfy the compatibility condition: $\a_v = \b_u$. We denote $|\a| = a$ and $|\b| = b$, and the first fundamental form can be rewritten as $\d s^2 = a^2 \d u^2 + 2 ab \cos \varphi \ \d u \ \d v + b^2\d v^2$, where $\vartheta$ the angle subtended by the vectors $\a, \b$. The normal to the surface along these network lines reduce to $\N = \a \times \b / (ab\sin \varphi)$.

Tchebychev nets are structures that describe a fabric, whose individual threads are inextensible i.e. the length of each individual element is preserved. This length constraint translates to $|\r_u| = |\r_v| = 1$ for every infinitesimal element. We find that the Tchebychev nets are a special case of NS materials whose all the edges are rigid for which we get $a = b = 1$. In order for the parameterization to represent a NS material, if the rigid links lie along lines of constant $v$, we require $a = 1$. In the ensuing subsections we find such a parameterisation for developable surfaces as well as elliptic and hyperbolic surfaces.

\subsection{Developable surfaces}
We know from geometry that any developable surface can be expressed as: $\r(u,v) = \al(v) + u \be(v)$ where $\al(v)$ is the directrix and $\be(v)$ is the unit vector pointing along the direction of ruling. We can immediately see that this form satisfies the requirement of NS material as $\r_u = \a = \be(v), |\beta(v)| = 1$ and further the stretching $b = (\alpha'(v))^2 + (u \beta'(v))^2 + 2 \alpha'(v) \beta'(v) \cos \vartheta$. As an example, let us consider a cylinder: $\r(u,v) = (\alpha \cos v, \alpha \sin v, 0) + u(0, 0, 1)$. We see that the direction along $u$ satisfies $a = |\r_u| = 1$, which implies that we have chosen this to be the direction of link and further this target shape is achieved from an initial grid of $(u, v)$. For this cylinder, we further have the stretching along $v$ given by $b = |\r_v| = \alpha$.

\subsection{Elliptical surfaces}
The NS material, as has been described earlier, can be described using any parametric representation which satisfied length constraint along $u$-parametrisation and is free along $v$. Thus for elliptic surface we can choose a representation along the principal curvatures in general. Here we take the example of a sphere for which the discrete simulations are presented in the main article. We can represent the NS material confined to a sphere of radius $r_o$ using $\r(u,v) = ( a(\alpha v) \sin (u/a(\alpha v)), \alpha v, a(\alpha v) \cos (u/a(\alpha v)))$ where $a(v) = \sqrt{r_o^2 - v^2}$. Again $\alpha$ is a degree of freedom and from this we can calculate $b = (r_o^4 \alpha^2 + (-r_o^2 + u^2) v^2 \alpha^4)/(r_o^2 - v^2 \alpha^2)^2$.

\subsection{Hyperbolic surfaces}
For the case of hyperbolic surfaces we can leverage the aysmptotic lines to be the directions along which NS structures lie. Along asymptotic lines of hyperbolic surfaces, $e = g = 0$ and the Gauss-Mainardi-Codazzi condition becomes
\begin{align}
\bigg( \frac{f}{H} \bigg)_u + 2 \Gamma^2_{12} \frac{f}{H} =& \ 0, \\
\bigg( \frac{f}{H} \bigg)_v + 2 \Gamma^1_{12} \frac{f}{H} =& \ 0,
\end{align}
where $H^2 = EG - F^2$. The gaussian curvature simplifies to $\K = -f^2/H^2 = -1/\varrho^2$. Now let $\varphi$ be the angle between the parametric lines, then $\cos \varphi = F/\sqrt{EG}, \sin \varphi = H/\sqrt{EG}$.
Without loss of generality, we can further rewrite $E = a^2 = \varrho^2 \ta^2, G = b^2 = \varrho^2 \tb^2$, Gauss-Mainardi-Codazzi condition becomes:
\begin{align}
\ta_v - \frac{1}{2} \frac{\varrho_v}{\varrho} \ta - \frac{1}{2} \frac{\varrho_u}{\varrho} \tb \cos \varphi =& 0, \\
\tb_u - \frac{1}{2} \frac{\varrho_u}{\varrho} \tb - \frac{1}{2} \frac{\varrho_v}{\varrho} \ta \cos \varphi =& 0,
\end{align}
and the expression for gaussian curvature becomes:
\begin{align}
\varphi_{uv} + \frac{1}{2} \bigg( \frac{\varrho_u}{\varrho} \frac{\tb}{\ta} \sin \varphi \bigg)_u - \frac{1}{2} \bigg( \frac{\varrho_v}{\varrho} \frac{\tb}{\ta} \sin \varphi \bigg)_v =& \ \ta\tb \sin \varphi.
\end{align}
Now, as we have done earlier, let us chose the direction of constant $v$ to be inextensible links. From this we have the constraint of $|\r_u| = a = 1$ or equivalently $\ta = 1/\varrho$, while the other asymptotic line need not satisfy the length constraint. The equations then reduce to
\begin{align}
\tb_u + \frac{\varrho_v}{\varrho^2} \bigg[ \frac{3}{4} \sec \varphi - \cos \varphi \bigg] =& 0, \\
\varphi_{uv} - \frac{3}{2} \bigg( \frac{\varrho_v}{\varrho} \tan \varphi \bigg)_u - \frac{1}{2} \bigg( \frac{\varrho_v}{\varrho} \sin \varphi \bigg)_v =& \ \frac{\tb}{\varrho} \sin \varphi.
\end{align}
\subsection{Tchebychev nets vs NS-nets}\label{sec:cheb}
NS-nets are closely related to Tchebychev nets when the inter-unit distance in NS structures is fixed. Tchebychev nets came about in the description of fabrics, in order to identify the right locations to introduce cuts in the fabric so that the tailored material fits a target surface. For the deformed shaped in Fig.~\ref{fig:suppl6}$(a)$, the angle $\phi$ corresponding to each unit cell agrees closely with the solutions for Tchebychev nets for the corresponding shapes when plotted as a function of $s = uv$ where $u$ is the parameterization along rigid rod and $v$ along ends connecting two NS structures. Note that in case of Tchebychev nets, the length in both $u$ and $v$ parametrization is preserved thus allowing only for in-plane shear. NS materials now extend this paradigm by allowing for large in-plane deformations since by assembly $v$ is allowed to change.

The description of memory-free materials using $b(u, v)$ and $\varphi(u,v)$ reduces to Tchebychev nets when $b(u, v) = 1$. For a K-surface (surface with constant negative Gaussian curvature) the equation for NS material reduces to $\varphi_{uv} = \tilde{f}(v) \sin \varphi/\varrho$, where $\tilde{f}(v)$ is an arbitrary function of $v$ arising out of the additional stretching degree of freedom that NS structures. However for Tchebychev nets, as the additional stretching degree of freedom is lost, these equations reduce to the famous sine-Gordan equation $\varphi_{uv} = \K \sin \varphi$ obtained by  setting $\ta = \tb = 1/\varrho$. Solution to this equation is evaluated for an hyperboloid, and sphere, shown in solid line in Fig.~\ref{fig:suppl6}$(b)$.

\clearpage
\section{Experimental methods}
\label{sec:materials}

\begin{figure}[h!]
\centering
\includegraphics[width = 0.9\textwidth]{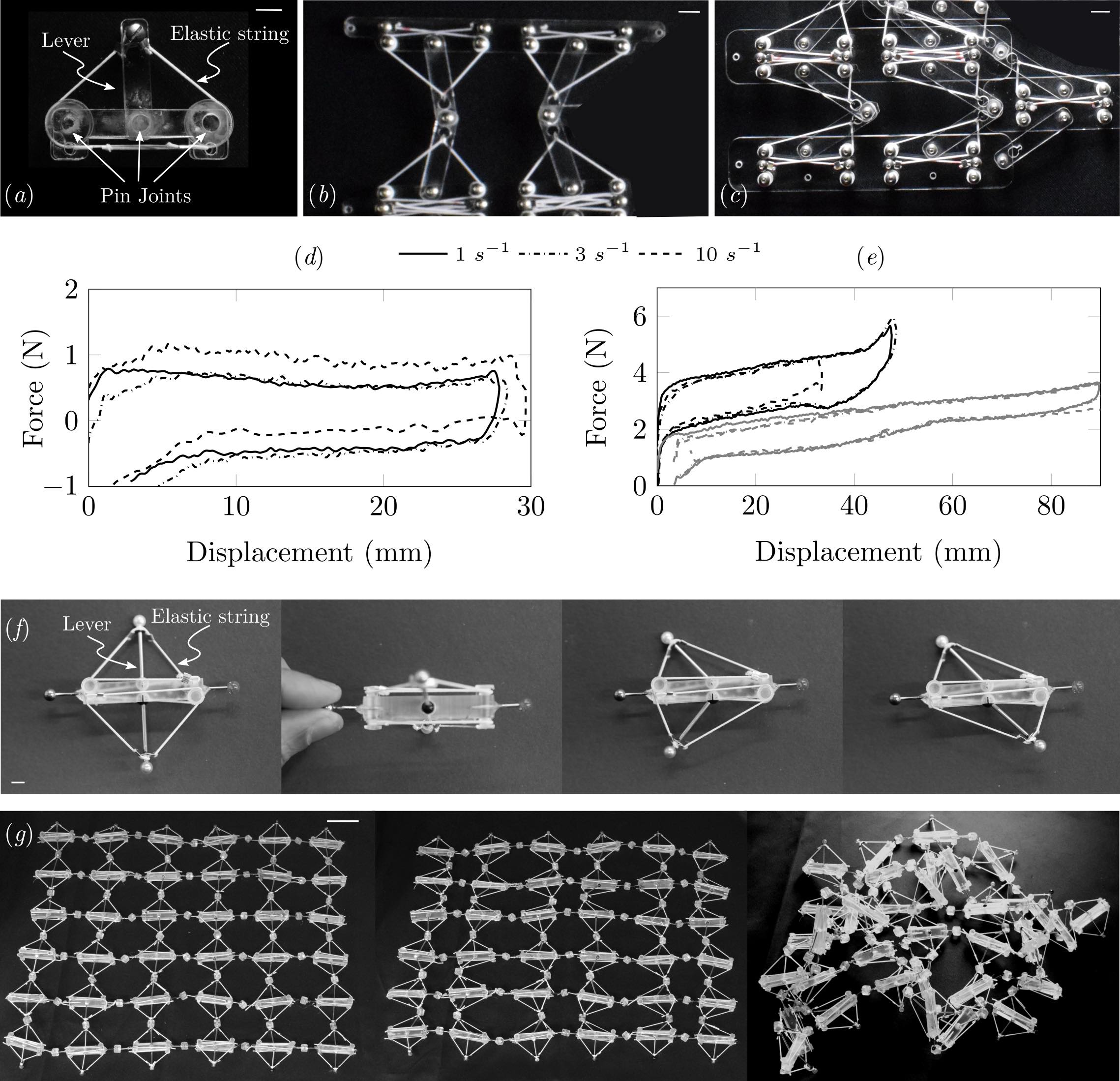}
\caption{$(a)$ Experimental realization of single unit of NS structure. Scale bar is 8 mm. Zoomed view of a $(b)$ negative Poisson structure, $(c)$ zero Poisson structure. Scale bar is 8 mm. $(d)$ Force vs displacement curves for single NS unit shown for different strain rates $1 s^{-1}, 3 s^{-1}, 10 s^{-1}$. $(e)$ Force displacement curve for the negative Poisson (black) and the zero Poisson (gray) structures for the same strain rates as in $(e)$. $(f)$ Front and top view of a unit of 3D NS cell and its different configurations. Scale bar is 6 mm. $(e)$ A sheet of NS material transforming from initial configuration to a state with all deformations on the lever angle and on the spherical joint. Scale bar is 4.5 cm.}
\label{fig:suppl4}
\end{figure}

\subsection{2D totimorphs fabrication}
The components of planar totimorphs are laser cut from a thin ($3$~mm) clear acrylic sheet. A planar totimorph has three components: a base link, a lever and an elastic band (see Fig.~\ref{fig:suppl3}$(a)$). The base link has two laser cut acrylic disks (that act like pulleys) that are sandwiched between two planar acrylic strips by pins. A single acrylic strip that acts as a lever is attached to the middle of a link by a pin joint whose length is twice that of the lever. To approximate zero-length springs, we use a prestretched elastic string whose natural physical length is ``hidden'' underneath the link (refer Fig.~\ref{fig:suppl4}$(a)$). The strings are then tied to the one end of the link and cut in lengths equal to the link length. The strings are then stretched over the pulley at the opposite end, and are then stretched to attach to the top end of the lever. This ensures that the effective stretch in each string is equal to the length between the link end and lever top. The negative and zero Poisson's ratio structures are constructed by an assembly of such unit cells resulting in planar totimorphs as shown in Fig.~3~$(c-d)$ and a zoomed view is shown in Fig.~\ref{fig:suppl4}$(b, c)$.

\subsection{NS-net fabrication}
We use a high resolution Stereolithography printer (Form 2 by Formlabs) to construct the 3D totimorphs. These are made from an assembly of unit cells shown in see Fig.~\ref{fig:suppl3}$(d)$ whose main components are the same as 2D totimorph. The key difference compared to the 2D unit cell is that the lever is attached to the link by a ball joint which enables out-of-plane displacement of the assembled structure. Multiple unit cells are assembled with ball joints between them. Thin metal rods of length 4 cm with ends made of spherical beads (6 mm diameter) are used as levers while the spheres at the end of the lever snap into the sockets in the link on one end and connects to other NS units through a pin-joint on the other end as shown in Fig.~\ref{fig:suppl5}$(a)$. This pin-joint is an approximation to a NS mechanism that couples angular deformation of the links to the deformation of the elastic strings.

\subsection{NS-net rigidification}
To demonstrate the shape morphing ability of the 3D structures (shown in Fig.~3$(e, f)$), we use reversible adhesive properties of Gallium (Sigma Aldrich 203319). A small drop ($\sim 500~\mu\text{l}$) of liquid Gallium is poured into the sockets, and the ball joint are then snapped in. At room temperature Gallium solidifies, and locks all the ball joints, hence locking the entire structure in place. On heating the structure to $40^{\circ}$C Gallium melts and the joints become free, and this enables the ability to morph the structure into any arbitrary shape.

\subsection{Spherical joints in experiments}
\begin{figure*}[h!]
\centering
\includegraphics[width=0.8\textwidth]{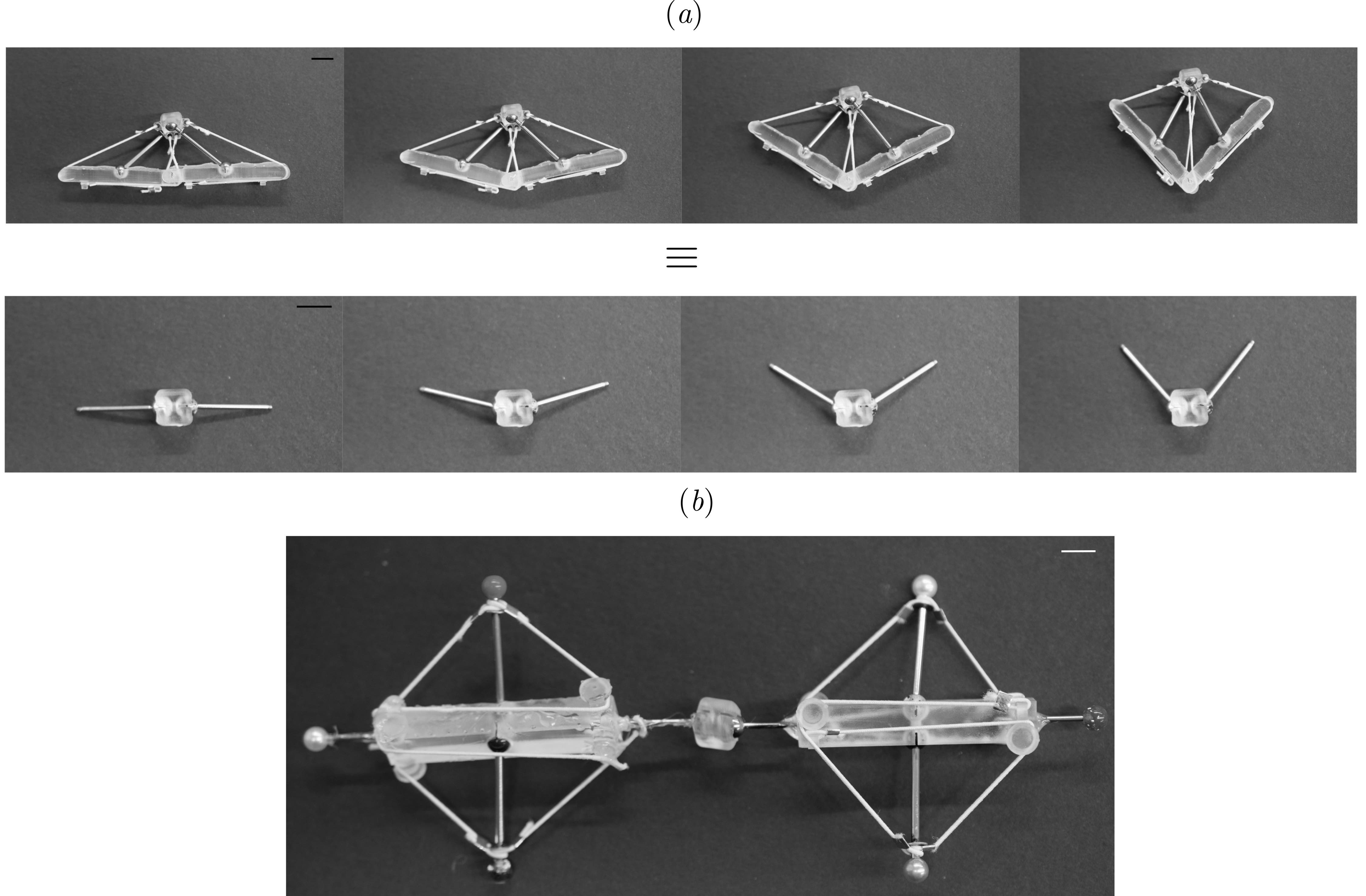}
\caption{$(a)$ \blue{Two neighboring NS unit cells in 3D are connected along the strut by planar joints shown on the top. However since the complexity of the fabrication procedure diverges, we approximate this joint using spherical joints, also used in the simulations. $(b)$ Experimental realization of the NS assembly with two unit cells connected using the approximated spherical joints}. Scale bar is 8 mm.}
\label{fig:suppl5}
\end{figure*}

\blue{As we have shown in the Fig.~2$(a)$, we can achieve different local curvatures using the constrained motion at different connections. Hereon we make the approximation that the joints connecting different levers and struts all are spherical joints as this simplifying assumption makes the computation as well as the experiments easier to handle. The primary purpose of the transformations shown in our simulations illustrate the efficacy of our algorithm to handle complex shapes while not compromising on the NS capabilities.}
In the 3D simulation further on as well as in experiments, we approximate this mini-link mechanism by a spherical joint shown in the lower panel of Fig.~\ref{fig:suppl5}$(a)$. The connection between two NS units in 3D used in experiments is shown in Fig.~\ref{fig:suppl5}$(b)$. The deformation field represented using angle $\varphi$ we have plotted in Fig.~\ref{fig:suppl6}$(b)$ shows that most of the strain in the structure is stored in the joints, which is why the deformation resembles that of a Tchebychev net. Thus it is useful to keep in mind that these deformations are in essence stored in the elastic energy of the strings attached to the mini-links detailed here. \blue{In Fig.~\ref{fig:suppl7}$(a)$ we show that the shape transforming capabilities of NS-nets are similar with our without the spherical joint approximation by comparing the solutions for fundamental shapes.}

\begin{figure*}[h!]
\centering
\includegraphics[width=0.6\textwidth]{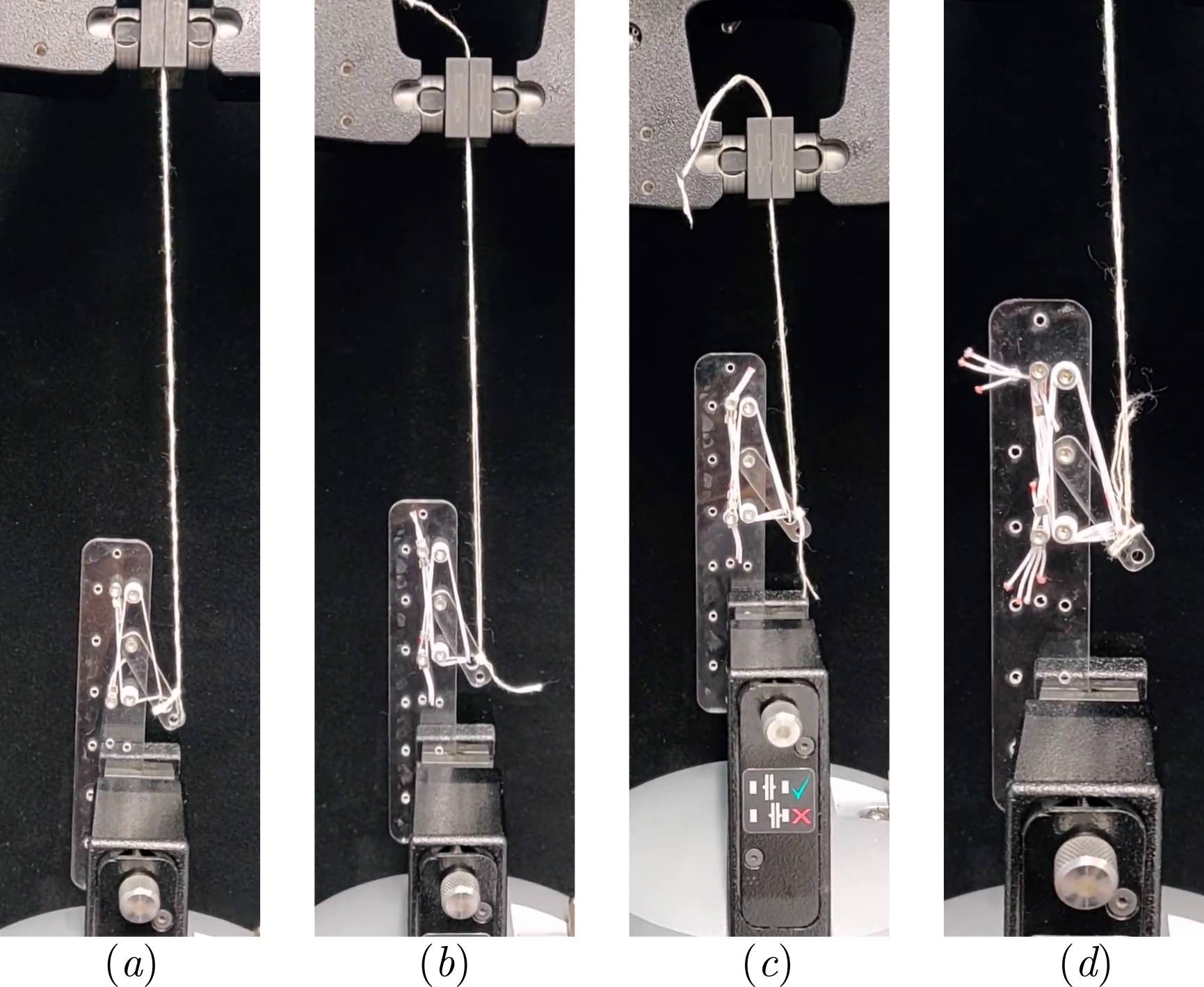}
\caption{Experimental protocol for data in Fig.~3$(b, c)$. The lever is pulled with an inextensible fiber connected to the measurement sensor. The measured force and displacement from the experiments, are converted to torque and orientation angle of the lever through simple geometric relations. Since, the fiber pulls the lever an angle which is generally not orthogonal to the direction of lever, a component of the pulling load also adds to the normal force at the pin joint, which creates deviations from a nearly flat plateau response in the torque. Note that when the load acts perpendicular to the orientation of the lever, an almost flat response in observed (Fig.~3$(c)$).}
\label{fig:suppl8}
\end{figure*}

\begin{figure*}[h!]
\centering
\includegraphics[width=0.6\textwidth]{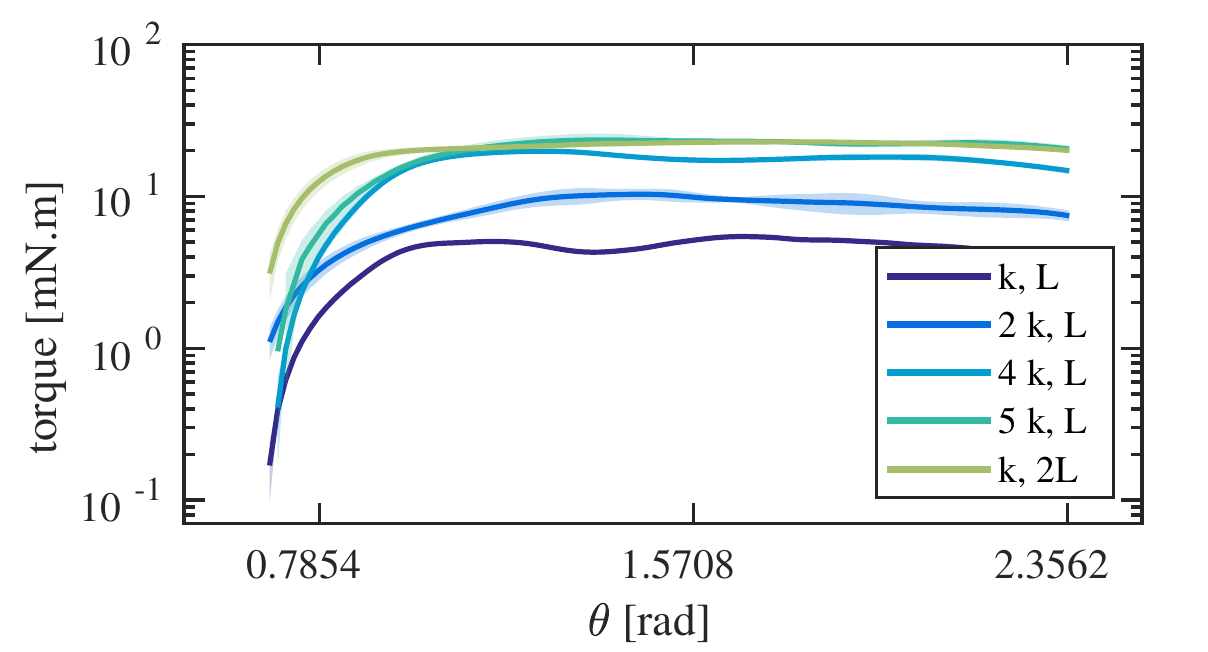}
\caption{Mechanical response of a fabricated unit cell shown in Fig.~3$(a)$. The torque required to change the angle $\vartheta$ in a unit cell. The torque has a plateau-like response over a wide range $\vartheta$, consistent with a (nearly) rigid perfectly-plastic material. The monotonic increase in the the torque at smaller angles is attributed to the experimental effects described in Fig.~\ref{fig:suppl8}.}
\label{fig:suppl9}
\end{figure*}

\section{List of videos}
\begin{itemize}
    \item Video 1: A Neutrally stable unit fabricated from a hidden-length elastic filament (an effective zero-length spring) and rigid rods, undergoes the angular deformation while being in equilibrium in all configurations.
    \item Video 2: A zero Poisson's ratio assembly constructed from unit neutrally stable units.
    \item Video 3: A negative Poisson's ratio assembly constructed from unit neutrally stable.
    \item Video 4: An ``on-demand'' NS-net that loses its rigidity when exposed to temperatures beyond threshold. The NS-net has Gallium filled joints, that drastically increase the friction in the joints when Gallium is its solid state, thus enabling a load-carrying capability. When the joints are heated with a heat gun, Gallium melts thus deceasing the friction between the joints, and structure collapses under the applied load.
    \blue{\item Video 5: A summary video with the highlights of totimorphic rigid plastic structural materials.}
\end{itemize}


\end{document}